%% file: RTPM.tex
\documentclass[runningheads]{llncs}
\usepackage[T1]{fontenc}

\usepackage{amsmath,amsfonts,amssymb}
\usepackage{graphicx}
\usepackage{textcomp}
\usepackage{xcolor}
\def\BibTeX{{\rm B\kern-.05em{\sc i\kern-.025em b}\kern-.08em
		T\kern-.1667em\lower.7ex\hbox{E}\kern-.125emX}}

\usepackage{etoolbox}
\usepackage{multirow}
\usepackage{tabularx}

\newcolumntype{C}{>{\centering\arraybackslash}X}
\usepackage{booktabs}
\usepackage{lipsum}

\usepackage{subcaption}
\usepackage{pgfplots}
\pgfplotsset{compat=1.16}
\newtheorem{lem}{Lemma}

\usepackage{algorithmic,float}
\usepackage[ruled,linesnumbered,noend]{algorithm2e}
\let\oldnl\nl
\newcommand{\nonl}{\renewcommand{\nl}{\let\nl\oldnl}}
\usepackage{caption} 
\makeatletter
\patchcmd\algocf@Vline{\vrule}{\vrule \kern-0.4pt}{}{}
\patchcmd\algocf@Vsline{\vrule}{\vrule \kern-0.4pt}{}{}
\makeatother

\SetKwInput{KwInput}{Input}                
\SetKwInput{KwOutput}{Output}              
\usepackage{tikz}
\usepackage{environ}
\newcommand\scalemath[2]{\scalebox{#1}{\mbox{\ensuremath{\displaystyle #2}}}}

\definecolor{ballblue}{rgb}{0.13, 0.67, 0.8}

\newlength\myindent
\makeatletter
\newcommand{\labeltext}[2]{%
	\@bsphack
	\csname phantomsection\endcsname 
	\def\@currentlabel{#1}{\label{#2}}%
	\@esphack
}
\makeatother
\newcommand*{\thead}[1]{\multicolumn{1}{|c|}{\bfseries #1}}

\usetikzlibrary{shapes.geometric,arrows,arrows.meta,calc,fit,matrix}
\definecolor{arrowcolor}{rgb}{.27,.45,.77}
\tikzstyle{arrow} = [arrowcolor,opacity=1,thin, {Triangle[angle=60:1.2mm]}-{Triangle[angle=60:1.2mm]}]
\tikzstyle{chart} = [rectangle, minimum width=3cm, minimum height=1cm, text centered,text width =3cm, draw=black, fill=white!30]
\tikzstyle{circlular} = [circle, minimum width=1cm, minimum height=1cm, text centered,text width =1cm, draw=black, fill=white!30]
\tikzstyle{circ} = [circle, minimum width=5mm, minimum height=5mm, text centered,text width =5mm, draw=black, fill=yellow!30]
\tikzstyle{roundrect3} = [rectangle,rounded corners, minimum width=3cm, minimum height=1.6cm, text centered,text width =1.4cm, draw=black, fill=red, text opacity=1]
\tikzstyle{roundrect4} = [rectangle,rounded corners, minimum width=3cm, minimum height=1.6cm, text centered,text width =1.4cm, draw=black, fill=blue, text opacity=1]
\tikzstyle{roundrect6} = [rectangle,rounded corners, minimum width=3cm, minimum height=1.6cm, text centered,text width =1.4cm, draw=black, fill=green, text opacity=1]
\tikzstyle{invisible} = [minimum width=3cm, minimum height=2cm, text centered,text width =1.4cm, draw=black, opacity = 1]
\tikzstyle{tinyarrow} = [thin,->,>=stealth]
\tikzstyle{strippedline} = [thick,dotted,>=stealth]
\BeforeBeginEnvironment{appendices}{\clearpage}

\makeatletter
\newcommand{\removelatexerror}{\let\@latex@error\@gobble}
\makeatother

\begin{document}
\title{Mining Rare Temporal Pattern in Time Series}
\author{Van Ho Long\inst{1,2} \and
Nguyen Ho\inst{3} \and
Trinh Le Cong\inst{4} \and
Anh-Vu Dinh-Duc\inst{1,2} \and
Tu Nguyen Ngoc\inst{5}}
\authorrunning{Van et al.}
\institute{School of Computer Science and Engineering, International University, Ho Chi Minh City, Vietnam \and
Vietnam National University, Ho Chi Minh City, Vietnam
\email{\{hlvan,ddavu\}@hcmiu.edu.vn}\\
 \and
Department of Computer Science, Loyola University Maryland, USA\\
\email{tnho@loyola.edu} \and
Department of Mathematics, Quy Nhon University, Vietnam\\
\email{lecongtrinh@qnu.edu.vn} \\
\and
TVT Group Corporation Research\\
\email{tu.nguyen@tvtgroup.io}}
\maketitle            
\vspace{-0.1in}
\begin{abstract}
Time series data from various domains is continuously growing, and extracting and analyzing temporal patterns within these series can provide valuable insights. Temporal pattern mining (TPM) extends traditional pattern mining by incorporating event time intervals into patterns, making them more expressive but also increasing the computational complexity in terms of time and space. One important type of temporal pattern is the rare temporal pattern (RTP), which occurs infrequently but with high confidence. Mining these rare patterns poses several challenges, for example, the low support threshold can lead to a combinatorial explosion and the generation of many irrelevant patterns.
To address this, an efficient approach to mine rare temporal patterns is essential. This paper introduces the Rare Temporal Pattern Mining from Time Series (RTPMfTS) method, designed to discover rare temporal patterns. The key contributions of this work are as follows: (1) An end-to-end RTPMfTS process that takes time series data as input and outputs rare temporal patterns. (2) A highly efficient Rare Temporal Pattern Mining (RTPM) algorithm, which leverages optimized data structures for fast event and pattern retrieval, as well as effective pruning techniques to accelerate the mining process. (3) A comprehensive experimental evaluation of RTPM, demonstrating that it outperforms the baseline in both runtime and memory efficiency.

\end{abstract}

\input{Introduction}
\input{RelatedWork}
\input{Preliminary}
\input{RTPMining}

\input{Experiment}
\input{Conclusion}

%
%
%
%
\bibliographystyle{splncs04}
\bibliography{references}
\clearpage
\appendix
\begin{appendix}
	\input{Appendix}
\end{appendix}
\end{document}

%% file: Introduction.tex
\vspace{-0.35in}
\section{Introduction}\vspace{-0.1in}
A growing volume of time series data is being collected from IoT sensors across various settings, including smart meters, smart plugs, smart appliances for electricity usage, and GPS-enabled smartphones for tracking movement. However, this raw data needs to be transformed into actionable insights to support decision-making and optimization. Patterns extracted from time series data, especially those that include temporal information, can offer valuable insights for these purposes.
A valuable pattern type, known as {\em rare temporal patterns (RTPs)}, is characterized by low support but high confidence. For instance, in the smart city domain, a rare pattern might be: ([8:00, 10:00] Snow $\succcurlyeq$ [8:15, 8:45] HighWind $\between$ [8:20, 9:00] HighPedestrianInjury) (meaning Snow contains HighWind and overlaps HighPedestrianInjury). This pattern suggests that the appearance of snow and strong wind correlates with an increase in traffic accidents within a particular time frame. Although this pattern is rare, it helps transportation coordinators warn about potential traffic accidents in a timely manner. 

\textit{Challenges of mining rare temporal patterns.} 
Extracting temporal patterns from time series data is significantly more complex than extracting sequential patterns. The inclusion of temporal information increases both computational workload and event relationship complexity, adding an exponential factor of O($3^{h^2}$) to the search space complexity, resulting in a total complexity of O($m^h3^{h^2}$) (proof in \cite{ho2024rareefficient}). Mining rare temporal patterns is even more challenging due to the low support threshold, causing a combinatorial explosion and producing many irrelevant patterns. Existing methods for mining rare itemsets \cite{li2022asso}, \cite{cui2022asso}, \cite{ji2021asso}, \cite{cai2021asso} and rare sequential patterns \cite{rahman2019sequential}, \cite{iqbal2019sequential}, \cite{samed2017sequential} do not account for temporal aspects. To address this, we propose a novel \textit{rare temporal pattern mining} approach, offering a robust framework to overcome these challenges.


\textit{Contributions.} 
In this paper, we present our comprehensive {\em Rare Temporal Pattern Mining from Time Series (RTPMfTS) approach} which solves the above challenges. 
Our key contributions are: 
(1) We present {\em end-to-end RTPMfTS process} that receives a set of time series as input, and produces {\em rare temporal patterns} as output.  
(2) We propose the {\em efficient Rare Temporal Pattern Mining (RTPM) algorithm} to mine {\em rare temporal patterns}. The novelties of RTPM are: a) the use of an efficient data structure, Hierarchical Hash Tables, to enable fast retrieval of events and patterns during the mining process; and b) pruning techniques based on the Apriori principle and the transitivity property of temporal relations to enable faster mining. 
(3) We perform extensive experiments on real-world datasets, showing that our RTPM significantly outperforms the baselines on both runtime and memory usage. 


%% file: RelatedWork.tex
\vspace{-0.22in}
\section{Related work}\label{sec:relatedwork}\vspace{-0.12in}
\textit{Temporal pattern mining:} Compared to sequential pattern mining, temporal pattern mining (TPM) is rather a new research topic. One of the first papers in this area is of Kam et al. that uses a hierarchical representation to manage temporal relations \cite{kam2000discovering}, and based on that mines temporal patterns. However, the approach in \cite{kam2000discovering} suffers from \textit{ambiguity} in relations between events  when presenting temporal relations. 
In \cite{moskovitch2015fast}, Moskovitch et al. design a TPM algorithm using the transitivity property of temporal relations. They use this property to generate candidates by inferring new relations between events. In comparison, we use the transitivity property for effective pruning. 
The most recent work addressing TPM is proposed by Lee et al \cite{lee2020z}. They present Z-Miner that improves the mining efficiency over existing methods by employing two data structures: a hierarchical hash-based structure called Z-Table for time-efficient candidate generation and support count, and Z-Arrangement, a structure to efficiently store event intervals in temporal patterns for efficient memory consumption. However, Z-Miner do not employ the transitivity property of temporal relations. Thus, it is less efficient in runtimes and memory usage when working with large datasets.

\textit{Rare pattern mining:} In recent years, there has been growing interest in identifying rare patterns that appear infrequently within a database. Techniques for discovering rare patterns in time series, often referred to as rare motifs, are discussed in \cite{wu2007mining}, \cite{gao2017motif}, \cite{begum2014raremotif}. However, since time series motifs are essentially repeated subsequences within the time series, these techniques cannot handle temporal events and are inadequate for mining rare temporal patterns. Another related approach involves rare association rules \cite{li2022asso}, \cite{cui2022asso}, \cite{ji2021asso}, \cite{cai2021asso}, \cite{ali2021asso}, \cite{bouasker2020asso}, \cite{borah2020asso}, \cite{fournier2020asso}, \cite{biswas2019asso}, \cite{piri2018asso}, which identify rare associations between items in the database. Nevertheless, these methods are limited to discovering rare association rules among itemsets and cannot address temporal events and the complex temporal relationships between them. Another area of research focuses on rare sequential patterns \cite{rahman2019sequential}, \cite{iqbal2019sequential}, \cite{samed2017sequential}, \cite{rahman2016sequential}, \cite{zhu2016sequential}, \cite{ou2016sequential}. However, these methods only account for sequential occurrences between events and cannot capture more complex relationships, such as overlaps or contains between temporal events. To the best of our knowledge, no existing research addresses the mining of rare temporal patterns, specifically targeting the rare occurrences of temporal patterns within a time series database.

%% file: Preliminary.tex
\section{Preliminaries}\label{sec:preliminary}

\vspace{-0.1in}
\subsection{Temporal Event and Temporal Relation}\vspace{-0.02in}
\textbf{Definition 3.1} (Time series) A \textit{time series} $X=  x_1, x_2, ..., x_n$ is a chronologically ordered sequence of data values that measure the same phenomenon during an observation time period.

A \textit{symbolic time series} $X_S$ of a time series $X$ uses a mapping function $f$$:$ $X$$\rightarrow$$\Sigma_{X}$ that maps each value $x_i \in X$ to a symbol $\omega \in \Sigma_{X}$, results in a sequence of symbols \cite{ho2022efficientvldb}. The \textit{symbol alphabet} $\Sigma_{X}$ is the finite set of symbols used to encode $X$.

 For example, let $X$=1.6, 1.2, 0.3, 0.0 be a time series of the energy usage. The symbolic representation of $X$ is: $X_S$=On,On,Off,Off (On:$x_i \geq 0.5$, Off:$x_i < 0.5$). 

\hspace{-0.2in}\textbf{Definition 3.2} (Symbolic database) 
Given a set of time series $\mathcal{X}=\{X_1,...,X_n\}$, the set of symbolic representations of the time series in $\mathcal{X}$ forms a \textit{symbolic database} $\mathcal{D_{\text{SYB}}}$.

An example of the symbolic database $\mathcal{D}_{\text{SYB}}$ is shown in Table \ref{tbl:SymbolDatabase}. There are $4$ time series representing the energy usage of $4$ electrical appliances: \{Stove, Toaster, Clothes Washer, Iron\}. For brevity, we name the appliances respectively as \{S, T, W, I\}. 
All appliances have the same alphabet $\Sigma$ = \{On, Off\}.

\input{table/tableSymbolDatabase_SmallerSpace.tex}
\input{table/tableSequenceDatabase_smallerSpace.tex}

\hspace{-0.2in}\textbf{Definition 3.3} (Temporal event) A \textit{temporal event} $E$ in a symbolic time series $X_S$ is a tuple $E = (\omega, T)$ where $\omega \in \Sigma_X$ is a symbol, and $T=\{[t_{s_i}, t_{e_i}]\}$ is the set of time intervals during which $X_S$ is associated with the symbol $\omega$. 

\textbf{Instance of a temporal event:} An \textit{instance} of the temporal event $E = (\omega, T)$ is a  tuple $e = (\omega, [t_{s_i},t_{e_i}])$. $e$ represents a single occurrence of $E$ during $[t_{s_i},t_{e_i}]$. We use the notation $E_{\triangleright e}$ to denote that event $E$ has an instance $e$. 

For example, consider the symbolic representation of $T$ in Table \ref{tbl:SymbolDatabase}. The temporal event \textit{``Toaster is On''} is: (TOn, \{[7:35, 7:40], [8:45, 8:50]\}).

\textbf{Relations between Temporal Events:} Let $E_i$ and $E_j$ be two temporal events, and $e_i=(\omega_i,[t_{s_i}, t_{e_i}])$ and $e_j=(\omega_j,[t_{s_j}, t_{e_j}])$ be their corresponding instances. We rely on the Allen's relations model  \cite{allen} and define three
basic temporal relations between $E_i$ and $E_j$: \textit{Follows}, \textit{Contains}, \textit{Overlaps}. Furthermore, we add the \textit{buffer} $\epsilon$ to the relation's endpoints for flexibility, and ensure that the relations are \textit{mutually exclusive} (proof is in the full paper \cite{ho2024rareefficient}). Table \ref{tbl:relations} illustrates the three temporal relations and their conditions, with $\epsilon > 0$ being the buffer size, and $d_o$ be the minimal overlapping duration between two instances in an Overlaps relation. 

\hspace{-0.2in}\textbf{Definition 3.4} (Temporal pattern) Let $\Re$$=$\{Follows, Contains, Overlaps\} be the set of temporal relations. A \textit{temporal pattern} $\scalemath{0.85}{P=<}$$\scalemath{0.8}{(r_{12}, E_{1},E_{2}),...,(r_{(n-1)(n)},E_{n-1},E_{n})}$$\scalemath{0.85}{>}$ is a list of triples $(r_{\textit{ij}}$,$E_{i}$,$E_{j})$, each representing a relation $r_{\textit{ij}} \in \Re$ between two events $E_i$ and $E_j$.

A temporal pattern that has $n$ events is called an $n$-event pattern. We use $E_i \in P$ to denote that the event $E_i$ occurs in $P$, and $P_1 \subseteq P$ to say that a pattern $P_1$ is a sub-pattern of $P$.

\vspace{-0.18in}
\subsection{Temporal Sequence Database}\vspace{-0.02in}
\textbf{Definition 3.5} (Temporal sequence)
A list of $n$ event instances $S$$=$$<$$e_1,...,e_i$,..., $e_n$$>$ forms a \textit{temporal sequence} if the instances are chronologically ordered by their start times. 
Moreover, $S$ has size $n$, denoted as $|S| = n$. 

\hspace{-0.2in}\textbf{Definition 3.6} (Temporal sequence database)
A set of temporal sequences forms a \textit{temporal sequence database} $\mathcal{D}_{\text{SEQ}}$ where each row $i$ contains a temporal sequence $S_i$. 

Table \ref{tbl:SequenceDatabase} shows the temporal sequence database $\mathcal{D}_{\text{SEQ}}$, created from the symbolic database $\mathcal{D}_{\text{SYB}}$ in Table \ref{tbl:SymbolDatabase}.

\hspace{-0.2in}\textbf{Definition 3.7} (Temporal sequence supports a pattern) Let $S$$=$$<$$e_1$,...,$e_i$,...,$e_n$$>$ be a temporal sequence. We say that $S$ \textit{supports} a temporal pattern $P$, denoted as $P \in S$, iff $|S| \ge 2$ $\wedge$ $\forall (r_{\textit{ij}},E_i,E_j) \in P, $ $\exists (e_l, e_m) \in S$ such that $r_{\textit{ij}}$ holds between $E_{i_{\triangleright e_l}}$ and $E_{j_{\triangleright e_m}}$.


In Table \ref{tbl:SequenceDatabase}, consider the temporal sequence $S$ at the second sequence, including the two event instances $e_1$=(SOn, [7:35, 7:45]), $e_2$$=$(TOn, [7:35, 7:40]). Here, $S$ supports a 2-event pattern $P=$$<$(Contains, SOn$_{\triangleright e_1}$, TOn$_{\triangleright e_2}$).

\vspace{-0.2in}
\subsection{Rare Temporal Pattern}\vspace{-0.02in}
Given a  temporal sequence database $\mathcal{D}_{\text{SEQ}}$, we want to find patterns that occur rarely but with high confidence in $\mathcal{D}_{\text{SEQ}}$. 
We use \textit{support} and \textit{confidence} \cite{omiecinski2003alternative} to measure the frequency and the likelihood of a pattern.

\hspace{-0.2in}\textbf{Definition 3.8} (Support of a temporal pattern) 
The \textit{support} of a pattern $P$ is the number of sequences $S \in \mathcal{D}_{\text{SEQ}}$ that support $P$. \vspace{-0.02in}
\begin{equation}
	\vspace{-0.05in}
	\small
	\textit{supp}(P) = \lvert \{S \in \mathcal{D}_{\text{SEQ}} \textit{ s.t. } P \in S\} \rvert 
	\label{eq:support3}
	\vspace{-0.02in}
\end{equation}


The support of a group of events $(E_1,..., E_n)$, denoted as $\textit{supp}(E_1,..., E_n)$, is defined similarly to that of a temporal pattern. 

\hspace{-0.2in}\textbf{Definition 3.9} (Confidence of a temporal pattern)
The \textit{confidence} of a temporal pattern $P$ in $\mathcal{D}_{\text{SEQ}}$ is the fraction between $\textit{supp}(P)$ and the support of its most frequent event: \vspace{-0.05in} 
\begin{equation}
	\vspace{-0.05in}
	\small
	\textit{conf}(P) = \frac{\textit{supp}(P)}{\max_{1 \leq k \leq |P|}\{\textit{supp}(E_k) \}} 
	\label{eq:confidence}
\end{equation}
where $E_k \in P$ is a temporal event. 
Since the denominator in Eq. \eqref{eq:confidence} is the maximum support of the events in $P$, the confidence computed in Eq. \eqref{eq:confidence} is the \textit{minimum confidence} of a pattern $P$ in $\mathcal{D}_{\text{SEQ}}$, 
which is also called the \textit{all-confidence} as in \cite{omiecinski2003alternative}. 

The confidence of an event pair $(E_i, E_j)$, denoted as $\textit{conf}(E_i, E_j)$, is defined similarly to that of a temporal pattern. 
The \textit{support} of event(s)/ pattern(s) defined in Eq. \eqref{eq:support3} follows the same intuition as the traditional support concept, indicating how frequently an event/ pattern occurs in a given database, the \textit{confidence} computed in Eq. \eqref{eq:confidence} instead represents the minimum likelihood of an event pair/ pattern, knowing the likelihood of its most frequent event.

\textbf{Rare Temporal Pattern Mining.} Given a set of univariate time series $\mathcal{X}=\{X_1,...,X_n\}$, let $\mathcal{D}_{\text{SEQ}}$ be the temporal sequence database obtained from ${\mathcal{X}}$, and $\sigma_{\min}$, $\sigma_{\max}$ and $\delta$ be the minimum support, the maximum support and the minimum confidence thresholds, respectively. The Rare Temporal Pattern Mining from Time Series (RTPMfTS) problem aims to find all temporal patterns $P$ in $\mathcal{D}_{\text{SEQ}}$ so that: $\sigma_{\min} \leq \textit{supp}(P) \leq \sigma_{\max}$ $\wedge$ $\textit{conf}(P) \geq \delta$.

%% file: table/tableSymbolDatabase_SmallerSpace.tex
\begin{table*}[!t]
	\begin{minipage}{1\linewidth}
		\caption{A Symbolic Database $\mathcal{D}_{\text{SYB}}$}
		\centering
		\resizebox{\textwidth}{1.2cm}{
			\renewcommand{\arraystretch}{1.5}
			\begin{tabular}{|c|ccccccc|ccccccc|ccccccc|ccccccc|}
				\hline
				\textbf{Time} & \textbf{7:00} & \textbf{7:05} & \textbf{7:10} & \textbf{7:15} & \textbf{7:20} & \textbf{7:25} & \textbf{7:30} & \textbf{7:35} & \textbf{7:40} & \textbf{7:45} & \textbf{7:50} & \textbf{7:55} & \textbf{8:00} & \textbf{8:05} & \textbf{8:10} & \textbf{8:15} & \textbf{8:20} & \textbf{8:25} & \textbf{8:30} & \textbf{8:35} & \textbf{8:40} & \textbf{8:45} & \textbf{8:50} & \textbf{8:55} & \textbf{9:00} & \textbf{9:05} & \textbf{9:10} & \textbf{9:15} \\ 
				\hline 
				\textbf{S} & \normalsize Off & \normalsize Off & \normalsize Off & \normalsize Off & \normalsize Off & \normalsize Off & \normalsize Off & \normalsize On & \normalsize On & \normalsize On & \normalsize Off & \normalsize Off & \normalsize Off & \normalsize Off & \normalsize Off & \normalsize Off & \normalsize Off & \normalsize Off & \normalsize Off & \normalsize Off & \normalsize Off & \normalsize Off & \normalsize Off & \normalsize Off & \normalsize Off & \normalsize Off & \normalsize Off & \normalsize Off   \\ 			
				\hline 
					\textbf{T} & \normalsize Off & \normalsize Off & \normalsize Off & \normalsize Off & \normalsize Off & \normalsize Off & \normalsize Off & \normalsize On & \normalsize On & \normalsize Off & \normalsize Off & \normalsize Off & \normalsize Off & \normalsize Off & \normalsize Off & \normalsize Off & \normalsize Off & \normalsize Off & \normalsize Off & \normalsize Off & \normalsize Off & \normalsize On & \normalsize On & \normalsize Off & \normalsize Off & \normalsize Off & \normalsize Off & \normalsize Off   \\ 			
				\hline 
					\textbf{W} & \normalsize Off & \normalsize Off & \normalsize Off & \normalsize Off & \normalsize Off & \normalsize Off & \normalsize Off & \normalsize On & \normalsize On & \normalsize Off & \normalsize Off & \normalsize Off & \normalsize Off & \normalsize Off & \normalsize Off & \normalsize Off & \normalsize Off & \normalsize Off & \normalsize Off & \normalsize Off & \normalsize Off & \normalsize On & \normalsize On & \normalsize On & \normalsize Off & \normalsize Off & \normalsize Off & \normalsize Off   \\ 			
				\hline 
					\textbf{I} & \normalsize On & \normalsize On & \normalsize On & \normalsize On & \normalsize On & \normalsize On & \normalsize On & \normalsize Off & \normalsize Off & \normalsize Off & \normalsize Off & \normalsize Off & \normalsize Off & \normalsize Off & \normalsize On & \normalsize On & \normalsize On & \normalsize On & \normalsize On & \normalsize On & \normalsize On & \normalsize Off & \normalsize Off & \normalsize Off & \normalsize Off & \normalsize Off & \normalsize Off & \normalsize Off   \\ 			
				\hline 
			\end{tabular}
		}
		\label{tbl:SymbolDatabase}
	\end{minipage}
	\vspace{-0.2in}
\end{table*}

%% file: table/tableSequenceDatabase_smallerSpace.tex
\begin{table*}
	\begin{minipage}{.58\textwidth}
		\caption{Temporal Relations between Events}
		\normalsize
		\resizebox{\textwidth}{2.5cm}{
			\begin{tabular}{|m{.51\columnwidth}| m{1\columnwidth}|}
				\hline
				Follows: 
				$E_{i_{\triangleright e_i}} \rightarrow E_{j_{\triangleright e_j}}$ &  \begin{tikzpicture}
					
					\draw (0,0) -- node[below]{\textbf{e$_{i}$}} ++(0.7,0);
					\filldraw (0,0) circle (2pt)node[above]{\small t$_{s_i}$} ;
					\filldraw (0.7,0) circle (2pt)node[above ]{\small t$_{e_i}{\pm \epsilon}$};
					
					\filldraw (0.9,0) circle (2pt)node[below right]{\small t$_{s_j}$};
					\filldraw (2.3,0) circle (2pt)node[below]{\small t$_{e_j}$};
					\draw (0.9,0) -- node[above]{\textbf{e$_{j}$}} ++(1.4,0);
					
					\draw (3.5,0) -- node[below]{\textbf{e$_{i}$}} ++(0.7,0);
					\filldraw (3.5,0) circle (2pt)node[above]{\small t$_{s_i}$} ;
					\filldraw (4.2,0) circle (2pt)node[above]{\small t$_{e_i}{\pm \epsilon}$};
					
					\filldraw (4.7,0) circle (2pt)node[below right]{\small t$_{s_j}$};
					\filldraw (6.1,0) circle (2pt)node[below]{\small t$_{e_j}$};
					\draw (4.7,0) -- node[above]{\textbf{e$_{j}$}} ++(1.4,0);
					
					\node [align=center] at (3.5,-0.7) {\small t$_{e_i}{\pm \epsilon}$ $\le$ t$_{s_j}$};
					
				\end{tikzpicture}  \\ \hline
				
				Contains: $E_{i_{\triangleright e_i}} \succcurlyeq E_{j_{\triangleright e_j}}$ & 	\begin{tikzpicture}
					\draw (0,0) -- node[above]{\textbf{e$_i$}} ++(2.0,0);
					\filldraw (0,0) circle (2pt)node[above]{\small t$_{s_i}$} ;
					\filldraw (2,0) circle (2pt)node[above]{\small t$_{e_i}\pm \epsilon$};
					
					\draw (0,-0.5) -- node[above]{\textbf{e$_{j}$}} ++(2.0,0);
					\filldraw (0,-0.5) circle (2pt)node[below]{\small t$_{s_j}$} ;
					\filldraw (2,-0.5) circle (2pt)node[below]{\small t$_{e_j}$};
										
					\draw (3.5,0) -- node[above]{\textbf{e$_i$}} ++(2.1,0);
					\filldraw (3.5,0) circle (2pt)node[above]{\small t$_{s_i}$} ;
					\filldraw (5.6,0) circle (2pt)node[above]{\small t$_{e_i}\pm \epsilon$};	
					
					\draw (3.9,-0.5) -- node[above]{\textbf{e$_j$}} ++(1.4,0);
					\filldraw (3.9,-0.5) circle (2pt)node[below]{\small t$_{s_j}$} ;
					\filldraw (5.3,-0.5) circle (2pt)node[below]{\small t$_{e_j}$};	
					
					\draw (0,-1.5) -- node[above]{\textbf{e$_i$}} ++(2.0,0);
					\filldraw (0,-1.5) circle (2pt)node[above]{\small t$_{s_i}$} ;
					\filldraw (2.0,-1.5) circle (2pt)node[above]{\small t$_{e_i}\pm \epsilon$};
					
					\draw (-0,-2) -- node[above]{\textbf{e$_{j}$}} ++(1.5,0);
					\filldraw (0,-2) circle (2pt)node[below]{\small t$_{s_j}$} ;
					\filldraw (1.5,-2) circle (2pt)node[below]{\small t$_{e_j}$};
					
					\draw (3.6,-1.5) -- node[above]{\textbf{e$_i$}} ++(2.0,0);
					\filldraw (3.6,-1.5) circle (2pt)node[above]{t$_{s_i}$} ;
					\filldraw (5.6,-1.5) circle (2pt)node[above]{t$_{e_i}\pm \epsilon$};
					
					\draw (4.1,-2) -- node[above]{\textbf{e$_{j}$}} ++(1.5,0);
					\filldraw (4.1,-2) circle (2pt)node[below]{t$_{s_j}$} ;
					\filldraw (5.6,-2) circle (2pt)node[below]{t$_{e_j}$};
					
					\node [align=center] at (3.0,-2.7) {\small (t$_{s_{i}} \le$ t$_{s_j}$)  $\wedge$ (t$_{e_i}{\pm \epsilon}$ $\ge$ t$_{e_j}$)};
					
				\end{tikzpicture} \\ \hline
				
				Overlaps: $E_{i_{\triangleright e_i}} \between E_{j_{\triangleright e_j}}$ & 
				 \begin{tikzpicture}
					
					\draw (0,0) -- node[above]{\textbf{e$_i$}} ++(2,0);
					\filldraw (0,0) circle (2pt)node[above]{\small t$_{s_i}$} ;
					\filldraw (2,0) circle (2pt)node[above]{\small t$_{e_i}\pm \epsilon$};
					
					\draw (1,-0.75) -- node[above right]{\textbf{e$_{j}$}} ++(2.0,0);
					\filldraw (1,-0.75) circle (2pt)node[below]{\small t$_{s_j}$} ;
					\filldraw (3,-0.75) circle (2pt)node[below]{\small t$_{e_j}$};
					
					\draw[dashed] (1,0) -- (1,-0.75);
					\draw[dashed] (2,0) -- (2,-0.75);
					
					\draw[dashed,>=latex,thin,<->] (1,-0.325) -- node[above]{d$_{o}$} ++(1,0);
					
					\node [align=center] at (1.75,-1.45) {\small (t$_{s_i}<$ t$_{s_j}$) $\wedge$ (t$_{e_i}{\pm \epsilon}$ $<$ t$_{e_j}$) $\wedge$ 
						\small (t$_{e_i}$ $-$ t$_{s_j}$ $\ge$ d${_o}{\pm \epsilon}$)};
					
				\end{tikzpicture} \\
				\hline
		\end{tabular} }
		\label{tbl:relations}
	\end{minipage}
	\begin{minipage}{.42\textwidth}
	\caption{\small A Temporal Seq. DB $\mathcal{D}_{\text{SEQ}}$}
	\label{tbl:SequenceDatabase}
	\resizebox{\textwidth}{2.5cm}{
		\begin{tabular}{ |c| m{4.6cm}| }
			\hline  \textbf{ID} & \textbf{\;\;\;\;\;\; Temporal sequences} \\
			\hline  
			1   & \scriptsize {(SOff,[7:00,7:30]),  (TOff,[7:00,7:30]),  (WOff,[7:00,7:30]),  (IOn,[7:00,7:30]) } 
			\\
			\hline 
			2   &  \scriptsize {(SOn,[7:35,7:45]), (TOn,[7:35,7:40]), (WOn,[7:35,7:40]), (IOff,[7:35,8:05]), (TOff,[7:40,8:05]), (WOff,[7:40,8:05]), (SOff,[7:45,8:05])
    }  
			\\ 
			\hline 
			3   &  \scriptsize {(SOff,[8:10,8:40]), (TOff,[8:10,8:40]), (WOff,[8:10,8:40]), (IOn,[8:10,8:40])}
			\\
			\hline 
			4   &  \scriptsize {(SOff,[8:45,9:15]), (TOn,[8:45,8:50]), (WOn,[8:45,8:55]), (IOff,[8:45,9:15]), (TOff,[8:50,9:15]), (WOff,[8:55,9:15])}
			\\
			\hline
		\end{tabular} 
	}
	\end{minipage}	
	\vspace{-0.3in}
\end{table*}

%% file: RTPMining.tex
\vspace{-0.2in}
\section{Rare Temporal Pattern Mining} \label{sec:FTPMfTSMining}
\vspace{-0.15in}
\subsection{Overview of RTPMfTS Mining Process}\vspace{-0.02in}
The RTPMfTS process consists of $2$ phases. The first phase, \textit{Data Transformation}, converts a set of time series $\mathcal{X}$ into a symbolic database $\mathcal{D}_{\text{SYB}}$, and then converts $\mathcal{D}_{\text{SYB}}$ into a temporal sequence database $\mathcal{D}_{\text{SEQ}}$. The second phase, \textit{Rare Temporal Pattern Mining (RTPM)}, mines rare patterns which includes $3$ steps: (1) \textit{Single Event Mining}, (2) \textit{Rare 2-Event Pattern Mining}, and (3) \textit{Rare k-Event Pattern Mining} (\text{k}$>$$2$). The final output is a set of all rare patterns in $\mathcal{D}_{\text{SEQ}}$.

\input{FTPMiningAlg}

%% file: FTPMiningAlg.tex
\vspace{-0.2in}
\subsection{Rare Temporal Pattern Mining}\vspace{-0.02in}
We now present our RTPM to mine rare temporal patterns from $\mathcal{D}_{\text{SEQ}}$. The main novelties of RTPM are: a) the use of the \textit{Hierarchical Lookup Hash} structure \cite{ho2022seasonal}, to enable fast retrieval of events and patterns during the mining process, and b) the proposal of two pruning techniques based on the Apriori principle and the temporal transitivity property. Algorithm \ref{algorithmHTPGM} provides the pseudo-code of RTPM.

\SetNlSty{}{}{:} 	
\begin{algorithm}[!t]
	\algsetup{linenosize=\tiny}
	\SetInd{0.5em}{0.5em}
	\DontPrintSemicolon
	
	\caption{\mbox{Rare Temporal Pattern Mining}}
	\label{algorithmHTPGM}
	
	\KwInput{Temporal sequence database $\mathcal{D_{\text{SEQ}}}$, minimum support threshold $\sigma_{\min}$, maximum support threshold $\sigma_{\max}$, minimum confidence threshold $\delta$}
	\KwOutput{The set of rare temporal patterns $P$}
	
	\ForEach{\textit{event} $E_i \in \mathcal{D_{\text{SEQ}}}$}{
		Compute $\textit{supp}(E_i)$;\;
		\If{$\textit{supp}(E_i) \geq \sigma_{min}$} {
			Insert $E_i$ to \textit{1Event};
		}
	}	
	
	\nonl //Mining rare 2-event patterns \;
	EventPairs $\leftarrow$ Cartesian(\textit{1Event},\textit{1Event});\;  
	FilteredPairs $\leftarrow$ $\emptyset$;\;
	\ForEach{$(E_i,E_j)$ in EventPairs}{
		Compute $\textit{supp}(E_i,E_j)$;\;
		\If{$\textit{supp}(E_i,E_j) \geq \sigma_{min}$}{
			FilteredPairs $\leftarrow$ Apply\_Lemma4($E_i,E_j$);\;
		}
	}
	\ForEach{$(E_i,E_j)$ in FilteredPairs}{
		Retrieve event instances and check relations against $\sigma_{\min}$, $\sigma_{\max}$, $\delta$;\;  
	}	
	
	\nonl //Mining rare k-event patterns \;
	Filtered1Freq $\leftarrow$ Transitivity\_Filtering(1Freq);\;
	kEvents $\leftarrow$ Cartesian(\textit{Filtered1Freq},\textit{(k-1)Freq});\;  
	FilteredkEvents $\leftarrow$ Apriori\_Filtering(kEvents);\;
	\ForEach{\textit{kEvents} in FilteredkEvents}{
		Retrieve relations and iteratively check relations against $\sigma_{\min}$, $\sigma_{\max}$, $\delta$;
	}	
\end{algorithm}

\input{1Frequent}

\begin{figure*}[!t]
	\setlength{\tabcolsep}{0pt}
	\begin{tabularx}{\linewidth}{ll}
		\begin{minipage}{.33\linewidth}
			\begin{minipage}{\linewidth}
				\captionsetup{justification=centering, font=small}
				\includegraphics[width=1\textwidth,height=3.1cm]{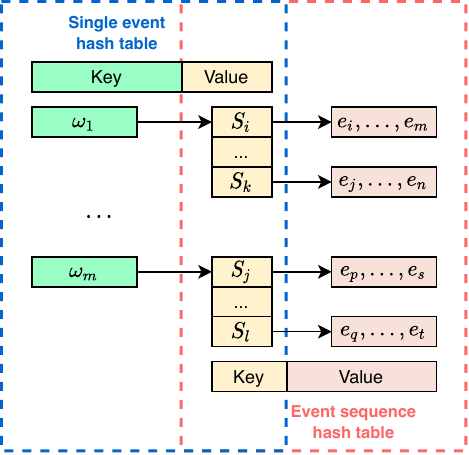}
				\vspace{-0.2in}
				\caption{The $HLH_1$ structure}
				\label{fig:hlh1}
			\end{minipage}
		\end{minipage}&
		\begin{minipage}{.66\linewidth}
			\begin{minipage}{\linewidth}
				\centering
				\captionsetup{justification=centering, font=small}
				\includegraphics[width=1\textwidth,height=3.1cm]{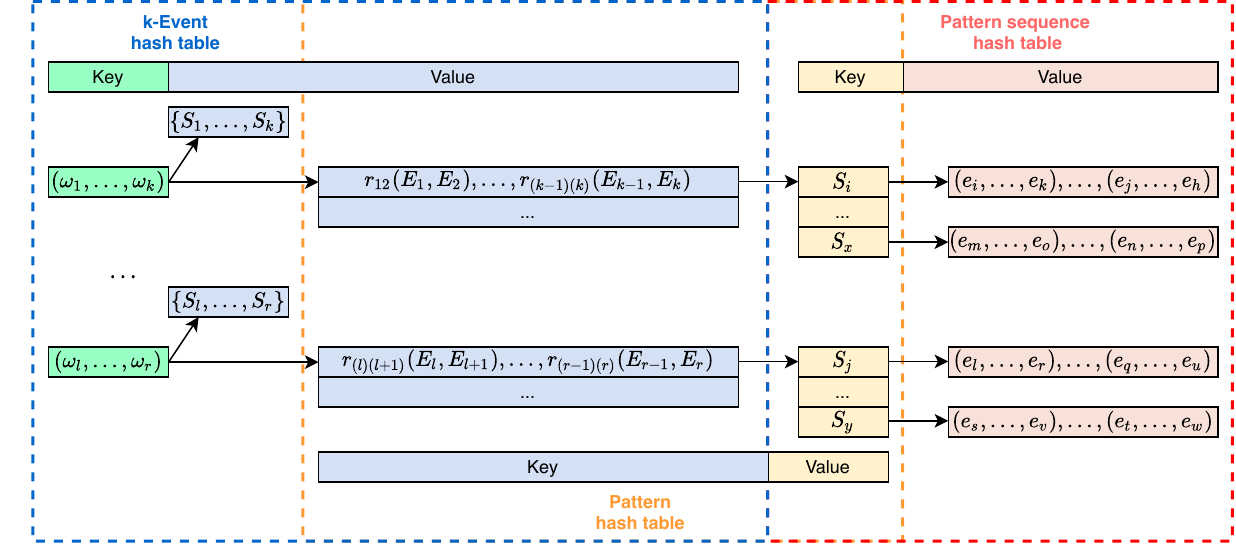}
				\vspace{-0.2in}
				\caption{The $HLH_k (k \geq 2)$ structure}
				\label{fig:hlhk}
			\end{minipage}
		\end{minipage}
	\end{tabularx}
\end{figure*}

\begin{figure}[!t]
	\centering
	\captionsetup{justification=centering, font=small}
	\begin{minipage}{\linewidth}
		\includegraphics[width=\textwidth,height=2.5cm]{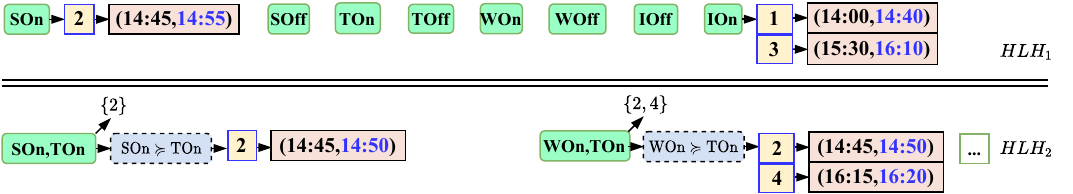}
		\vspace{-0.2in}
		\caption{A hierarchical lookup hash tables for the running example}
		\label{fig:patternTree}
	\end{minipage}
\vspace{-0.3in}
\end{figure}

\input{2Frequent}

\input{kFrequent}

%% file: 1Frequent.tex
\vspace{-0.1in}
\subsection{Mining Single Events}\vspace{-0.02in} \label{sec:1freq}
\textbf{Hierarchical lookup hash structure $HLH_1$:} Fig. \ref{fig:hlh1} illustrates the hierarchical lookup hash structure $HLH_1$. $HLH_1$  consists of two hash tables: the \textit{single event hash table} $EH$, 
and the \textit{event sequence hash table} $SH$. 
Each hash table has a list of $<$key, value$>$ pairs. 
In $EH$, the key is the event symbol $\omega \in \Sigma_X$ representing the event $E_i$, and the value is the set of sequences where $E_i$ occurs, arranged in an increasing order. In $SH$, the key is taken from the value component of $EH$, while the value stores event instances of $E_i$ that appear at the corresponding sequence in $\mathcal{D}_{\text{SEQ}}$.
The $HLH_1$ structure enables fast retrieval of event sequences and instances when mining k-event patterns. 

\textbf{Mining Single Events:} 
The first step in RTPM is to find single events (Alg. \ref{algorithmHTPGM}, lines 1-4). For each event $E_i$, the support \textit{supp}($E_i$) is  computed by first scanning $\mathcal{D}_{\text{SEQ}}$, and determine if \textit{supp}($E_i$)$\geq \sigma_{\min}$, then $E_i$ is inserted into \textit{1Event}. Note that for single events, we do not consider the constraints on the \textit{confidence}, since confidence of single events is always $1$ and on maximum support $\sigma_{\max}$ because of the following lemma. 

\begin{lem}\label{lem0}\vspace{-0.05in}
	Let $P$ be a temporal pattern and $E_i$ be a single event such that $E_i \in P$. Then $\textit{supp}(P) \leq \textit{supp}(E_i)$.
	\vspace{-0.05in}
\end{lem}

\textbf{Proof.} \textit{Detailed proofs of all lemmas and complexities in this article can be found in the Appendix of the full paper \cite{ho2024rareefficient}.}

From Lemma \ref{lem0}, a single event $E_i$ whose support \textit{supp}$(E_i) > \sigma_{\max}$ can form a pattern $P$ that has $\textit{supp}(P) \le \sigma_{\max}$. Thus, the constraint on $\sigma_{\max}$ is not considered for single events to avoid the loss of potential temporal patterns.

We provide an example of $HLH_1$ in Fig. \ref{fig:patternTree} using data in Table \ref{tbl:SequenceDatabase}, with $\sigma_{\min}=0.2$. Here, we have $8$ events in $HLH_1$. 


%% file: 2Frequent.tex
\vspace{-0.1in}
\subsection{Mining Rare  2-event Patterns}\label{sec:2freq} \vspace{-0.02in}


\textbf{The hierarchical lookup hash structure $HLH_k$:} Fig. \ref{fig:hlhk} illustrates the \textit{hierarchical lookup hash structure} $HLH_k$ $(k \geq 2)$ to maintain k-event groups and patterns. 
The $HLH_k$ contains three hash tables: the \textit{k-event hash table} $EH_k$, the \textit{pattern hash table} $PH_k$, and the \textit{pattern sequence hash table} $SH_k$. Each hash table has a list of $<$key, value$>$ pairs.
In $EH_k$, {\em key} is the list of symbols $(\omega_1 ..., \omega_k)$ representing the k-event group $(E_1,...,E_k)$, and {\em value} is an \textit{object} which consists of two components: (1) the list of sequences where $(E_1,...,E_k)$ occurs (arranged in increasing order), and (2) a list of k-event temporal patterns  
created from the k-event group $(E_1,...,E_k)$.  
In $PH_k$, {\em key} is the pattern $P$ which takes the {\em value} component of $EH_k$, while {\em value} is the list of sequences that support $P$.  
In $SH_k$, {\em key} is the list of sequences supporting $P$ which takes the {\em value} component of $PH_k$, while {\em value} is the list of event instances from which the temporal relations in $P$ are formed. 
The $HLH_k$ hash structure helps speed up the k-event group mining through the use of the list of sequences in $EH_k$, and enables fast search for temporal relations between $k$ events using the information in $PH_k$ and $SH_k$.

\vspace{-0.2in}\subsubsection{Two-steps filtering approach}
Given the huge set of pattern candidates, it is expensive to check their support and confidence. We propose a \textit{filtering approach} to reduce the unnecessary candidate checking. Specifically, the mining process is divided into two steps: (1) it first finds k-event groups that satisfy the $\sigma_{\min}$ and $\delta$ constraints, (2) it then generates temporal patterns from those k-event groups. The correctness of this filtering approach is based on the Apriori-inspired lemmas below.
\begin{lem}\label{lem2}\vspace{-0.05in}
	Let $P$ be a 2-event pattern formed by an event pair $(E_i, E_j)$. Then, $\textit{supp}(P) \le \textit{supp}(E_i,E_j)$. 
	\vspace{-0.05in}
\end{lem} 

From Lemma \ref{lem2}, the support of a pattern is at most the support of its events. Thus, event pairs that do not satisfy the minimum support $\sigma_{\min}$ cannot create patterns satisfying $\sigma_{\min}$ and thereby, can be safely pruned.

\begin{lem}\label{lem3}	\vspace{-0.05in}
	Let $(E_i, E_j)$ be a pair of events occurring in a 2-event pattern $P$.  Then \textit{conf}($P$) $\le$ \textit{conf}($E_i,E_j$).
\vspace{-0.05in}
\end{lem}

From Lemma \ref{lem3}, the confidence of a pattern $P$ is always at most the confidence of its events. Thus, a low-confidence event pair cannot form any high-confidence patterns and therefore, can be safely pruned.
Applying Lemmas \ref{lem2} and \ref{lem3} to the first filtering step will remove low-support or low-confidence event pairs, reducing the candidate patterns of RTPM. Moreover, the constraint on $\sigma_{\max}$ is not considered in this filtering step to avoid the loss of 2-event patterns, since event pairs that do not satisfy the $\sigma_{\max}$ constraint can still form 2-event patterns satisfying $\sigma_{\max}$ (Lemma \ref{lem2}).

\textbf{Step 2.1. Mining event pairs considering $\sigma_{\min}$ and $\delta$:}
This step finds event pairs in $\mathcal{D}_{\text{SEQ}}$ satisfying $\sigma_{\min}$ and $\delta$, using the set \textit{1Event} found in $HLH_1$  (Alg. \ref{algorithmHTPGM}, lines 5-10). 
First, RTPM generates all possible event pairs by calculating the Cartesian product \textit{1Event} $\times$ \textit{1Event}. Next, for each pair $(E_i, E_j)$, the set $\mathcal{S}_{ij}$ (representing the set of sequences where both events occur) is computed by taking the \textit{intersection} between the set of sequences $\mathcal{S}_i$ of $E_i$ in $HLH_1$ and the set of sequences $\mathcal{S}_j$ of $E_j$ in $HLH_1$. Finally, we compute the support $\textit{supp}(E_i, E_j)$, and comparing against $\sigma_{\min}$. 
If $\textit{supp}(E_i, E_j) \ge \sigma_{\min}$, $(E_i, E_j)$ is further filtered using Lemma \ref{lem3}: $(E_i, E_j)$ is selected only if its confidence is at least $\delta$. After this step, only event pairs satisfying $\sigma_{\min}$ and $\delta$ are kept in $EH_2$ of $HLH_2$. 

\textbf{Step 2.2. Mining rare 2-event patterns:} This step finds 2-event patterns from the event pairs found in step 2.1 (Alg. \ref{algorithmHTPGM}, lines 11-12), considering three constraints $\sigma_{\min}$, $\sigma_{\max}$, and $\delta$. For each event pair $(E_i,E_j)$, we use the set of sequences $\mathcal{S}_{ij}$ to check relations between $E_i$ and $E_j$.
Specifically, for each sequence $S \in \mathcal{S}_{ij}$, the pairs of event instances $(e_i,e_j)$ are extracted. 
The support and confidence of each relation $r(E_{i_{\triangleright e_i}},E_{j_{\triangleright e_j}})$ are compared against $\sigma_{\min}$ and $\delta$ thresholds, after which only relations satisfying these two constraints are selected and stored in $PH_2$, while their event instances are stored in $SH_2$. Examples of the relations in $HLH_2$ can be seen in Fig. \ref{fig:patternTree}, e.g., event pair (SOn, TOn). Next, we check the satisfaction of each relation $r(E_{i_{\triangleright e_i}},E_{j_{\triangleright e_j}})$ against $\sigma_{max}$. If it satisfies $\sigma_{max}$, then it is a rare temporal pattern. We also emphasize that $HLH_2$ only stores patterns that satisfy the two constraints $\sigma_{\min}$ and $\delta$.


%% file: kFrequent.tex
\vspace{-0.2in}
\subsection{Mining Rare k-event Patterns}\vspace{-0.02in}
Mining rare k-event patterns ($k \ge 3$) follows a similar process as rare 2-event patterns, with additional prunings based on the transitivity property of temporal relations.     

\textbf{Step 3.1. Mining k-event combinations considering $\sigma_{\min}$ and $\delta$} (Alg. \ref{algorithmHTPGM}, lines 13-15). 
Let \textit{(k-1)Event} be the set of (k-1)-event combinations found in $HLH_{k-1}$, and \textit{1Event} be the set of single events in $HLH_1$. To generate all k-event combinations, the typical process is to compute the Cartesian product: \textit{(k-1)Event} $\times$ \textit{1Event}. However, using \textit{1Event} to generate k-event combinations at $HLH_k$ can create redundancy, since \textit{1Event} might contain events that when combined with \textit{(k-1)Event}, the combinations cannot form any patterns satisfying the $\sigma_{\min}$ constraint. For example, consider the event IOn at $HLH_1$ in Fig. \ref{fig:patternTree}. Here, IOn can be combined with event pairs in $HLH_2$ such as (SOn, TOn) to create a 3-event combination (SOn, TOn, IOn). However, (SOn, TOn, IOn) cannot form any 3-event patterns whose support is greater than $\sigma_{\min}$, since IOn is not present in any 2-event patterns in $HLH_2$. The combination (SOn, TOn, IOn) should not be created in the first place. We rely on the \textit{transitivity property} of temporal relations to identify such event combinations.  

\begin{lem} \label{lem:transitivity}\vspace{-0.05in}
Let $S=<e_1$,..., $e_{n-1}>$ be a temporal sequence that supports an (n-1)-event pattern $\scalemath{0.9}{P=<(r_{12}, E_{1_{\triangleright e_1}}, E_{2_{\triangleright e_2}}),..., (r_{(n-2)(n-1)}, E_{{n-2}_{\triangleright e_{n-2}}}, E_{{n-1}_{\triangleright e_{n-1}}})>}$. Let $e_n$ be a new event instance added to $S$ to create the temporal sequence $S^{'}$$=$$<e_1, ..., e_{n}>$. 
The set of temporal relations $\Re$ is transitive on $S^{'}$: $\forall e_i \in S^{'}$, $i < n$, $\exists r \in \Re$ s.t. $r(E_{i_{\triangleright e_i}}$,$E_{n_{\triangleright e_n}})$ holds.
		
\vspace{-0.05in}
\end{lem}

Lemma \ref{lem:transitivity} says that given a temporal sequence $S$, a new event instance added to $S$ will form at least one temporal relation with existing instances in $S$. 

\begin{lem}\label{lem:filter}\vspace{-0.05in}
	Let $N_{k-1}=(E_1,...,E_{k-1})$ be a (\textit{k-1})-event combination, and $E_k$ be a single event. The combination $N_k= N_{k-1} \cup E_k$ can form k-event temporal patterns whose support is at least $\sigma_{\min}$ if $\forall E_i \in N_{k-1}$, $\exists r \in \Re$ s.t. $r(E_i,E_k)$ is a temporal relation satisfying $\sigma_{\min}$. 
	\vspace{-0.05in}
\end{lem}
From Lemma \ref{lem:filter}, only single events in $HLH_1$ that occur in $HLH_{k-1}$ should be used to create k-event combinations.
Using this result, a filtering on \textit{1Event} is performed before calculating the Cartesian product. Specifically, from the events in $HLH_{k-1}$, we extract the distinct single events $D_{k-1}$, and \textit{intersect} them with \textit{1Event} to remove redundant single events: \textit{Filtered1Event} = $D_{k-1}$ $\cap$ \textit{1Event}. 
Next, the Cartesian product \textit{(k-1)Event} $\times$ \textit{Filtered1Event} is calculated to generate k-event combinations. Finally, we apply Lemmas \ref{lem2} and \ref{lem3} to select k-event combinations \textit{kEvent} which upheld the $\sigma_{\min}$ and $\delta$ constraints.

\textbf{Step 3.2 Mining rare k-event patterns:} This step finds rare k-event patterns (Alg. \ref{algorithmHTPGM}, lines 16-17). Unlike 2-event patterns, determining the relations in a k-event combination ($k \ge 3$) is much more expensive, as it requires to verify the frequency of $\frac{1}{2}k(k-1)$ triples. To reduce the cost of relation checking, we propose an iterative verification method that relies on the \textit{transitivity property} and the Apriori principle. 

\begin{lem}\label{lem5}\vspace{-0.05in}
	Let $P$ and $P^{'}$ be two temporal patterns. If $P^{'} \subseteq P$, then \textit{conf}($P^{'}$) $ \geq$ \textit{conf}($P$).
	\vspace{-0.05in}
\end{lem} 

\begin{lem} \label{lem6}\vspace{-0.05in}
Let $P$ and $P^{'}$ be two temporal patterns. If $P^{'} \subseteq P$ and \\ $\frac{\textit{supp}(P^{'})}{\max_{1 \leq k \leq \mid P \mid}\{\textit{supp}(E_k)\}}_{E_k \in P} \leq \delta$, then \textit{conf($P$)} $\leq \delta$.
\vspace{-0.05in}
\end{lem}

Lemma \ref{lem5} says that, the confidence of a pattern $P$ is always at most the confidence of its sub-patterns. Consequently, from Lemma \ref{lem6}, a temporal pattern $P$ cannot be high-confidence if any of its sub-patterns are low-confidence.

Let $N_{k-1}=(E_1,...,E_{k-1})$ be a (k-1)-event combination in $H_{k-1}$, $N_1=(E_k)$ be an event in $HLH_1$, and $N_k=N_{k-1} \cup N_1 = (E_1,...,E_k)$ be a k-event combination in $HLH_k$. To find k-event patterns for $N_k$, we first retrieve the set $P_{k-1}$ containing (k-1)-event patterns of $N_{k-1}$. 
Each $p_{k-1} \in P_{k-1}$ is a list of $\frac{1}{2}(k-1)(k-2)$ triples: $\{(r_{12}$, $E_{1_{\triangleright e_1}}$, $E_{2_{\triangleright e_2}})$,...,$(r_{(k-2)(k-1)}$, $E_{{k-2}_{\triangleright e_{k-2}}}$, $E_{{k-1}_{\triangleright e_{k-1}}})\}$. We iteratively verify the possibility of $p_{k-1}$ forming a k-event pattern with $E_k$ as follows. 
We first check whether the triple $(r_{(k-1)k}$, $E_{{k-1}_{\triangleright e_{k-1}}}$, $E_{k_{\triangleright e_{k}}})$ satisfies the constraints of $\sigma_{\min}$ and $\delta$. If the triple does not satisfy the $\sigma_{\min}$ (using Lemmas \ref{lem:transitivity} and \ref{lem:filter}) or $\delta$ (using Lemmas \ref{lem:transitivity}, \ref{lem5}, and \ref{lem6}), the verifying process stops immediately for $p_{k-1}$.  
Otherwise, we continue the verification as follows. Let $N^{'}_{k-1}=(E_1,...,E_{k-2}, E_k)$ be another (k-1)-event combination in $HLH_{k-1}$. We retrieve the set of frequent (k-1)-event patterns $P^{'}_{k-1}$ of $N^{'}_{k-1}$ such that $p^{'}_{k-1} \in P^{'}_{k-1}$ contains $\{(r_{12}, E_{1_{\triangleright e_1}}, E_{2_{\triangleright e_2}})$,...,$(r_{(k-3)(k-2)}, E_{{k-3}_{\triangleright e_{k-3}}}, E_{{k-2}_{\triangleright e_{k-2}}})\}$ $\in p_{k-1}$. If $p^{'}_{k-1}$ is not frequent or high-confidence, $p^{'}_{k-1}$ cannot form a frequent k-event pattern with $E_k$. Otherwise, we have a new k-event pattern formed between $p_{k-1}$ and $E_k$. 
Finally, to determine rare k-event patterns, we select
only k-event patterns in $PH_k$ that satisfy the constraint  $\sigma_{\max}$.



%% file: Experiment.tex
\vspace{-0.2in}
\section{Experimental Evaluation}\label{sec:experiment}\vspace{-0.1in}

Due to space limitations, we only present here the most important results, and discuss other findings in the Appendix of the full paper \cite{ho2024rareefficient}.
\vspace{-0.2in}
\subsection{Experimental Setup}\vspace{-0.1in}
\hspace{0.17in}\textbf{Datasets:}
We use three real-world datasets from three application domains: smart energy, smart city, and health. For smart energy (SE), we use energy data from electrical appliances in residential households \cite{nist}. For the \textit{smart city} (SC), we use weather and vehicle collision data obtained from NYC Open Data Portal \cite{smartcity}. For \textit{health}, we combine the \textit{influenza} (INF) dataset \cite{diseasedata} and weather data \cite{openweather} from Kawasaki, Japan. 

\textbf{Baseline methods:} Since our work is the first that studies rare temporal pattern mining, there is not an exact baseline to compare against RTPM. However, we adapt the state-of-the-art method for frequent temporal pattern mining Z-Miner \cite{lee2020z} to find rare temporal patterns. Specifically, for single events, we follow the original Z-Miner approach using the $\sigma_{\min}$. When mining k-event patterns ($k \ge 2$), we apply the following criteria to identify rare patterns: $\sigma_{\min}$, $\sigma_{\max}$ and $\delta$. The Adapted Rare Z-Miner is referred to as ARZ-Miner.

\textbf{Infrastructure:} We use a VM with 32 AMD EPYC cores (2GHz), 64 GB RAM, and 1 TB storage. 

\vspace{-0.2in}
\subsection{Qualitative Evaluation} \vspace{-0.05in}
\input{qualitativeEvaluation}
\vspace{-0.1in}
\input{rare_quantitativeEvaluation}

%% file: qualitativeEvaluation.tex
Table \ref{tbl:rarePatterns} shows several interesting rare temporal patterns extracted. Patterns P1-P4 are from SC and P5-P7 are from INF. Analyzing these patterns can reveal some rare but interesting relations between temporal events. For example, P1-P4 show there exists an association between extreme weather conditions and high accident numbers, such as high pedestrian injury during a heavy snowing day, which is very important to act on even though it occurs rarely. 

\input{table/tableInterestingPatterns}

%% file: table/tableInterestingPatterns.tex
\begin{table*}[!t]	
	\caption{Summary of Interesting Rare Patterns}
	\centering
	\resizebox{0.98\textwidth}{!}{
		\begin{tabular}{ |p{13.5cm}|c|c|c| }
			\hline  {\bfseries \;\;\;\;\;\;\;\;\;\;\;\;\;\;\;\;\;\;\;\;\;\;\;\;\;\;\;\;\;\;\;\;\;\;\;\;\;\;\;\;\;\;\;\;\;\;\;\;\;\;\;\;\;\;\;\;\;\;\; Patterns} & {\bfseries $\sigma_{\min}$ (\%)} & {\bfseries $\delta$ (\%)} & {\bfseries $\sigma_{\max}$ (\%)}\\	\hline  
			(P1) Heavy Rain $\succcurlyeq$ Unclear Visibility $\succcurlyeq$ Overcast Cloudiness $\rightarrow$ High Motorist Injury   &  5 & 30 & 9
			\\ \hline 
			(P2) Heavy Rain $\between$ Strong Wind $\rightarrow$ High Motorist Injury   &  2& 40 & 6 \\
			\hline
			(P3) Strong Wind $\between$ High Pedestrian Injury & 4 & 30 & 8\\
			\hline
			(P4) Extremely Unclear Visibility $\succcurlyeq$ High Snow $\succcurlyeq$ High Pedestrian Injury   &  3 & 45 & 7 
			\\ \specialrule{1.5pt}{1pt}{1pt}
			(P5) Frost Temperature $\between$ High Snow $\succcurlyeq$ High Influenza & 1 & 42 & 6
			\\ \hline 
			(P6) Low Temperature $\succcurlyeq$High Influenza & 1 & 42 & 6
			\\ \hline 
			(P7) Heavy Rain $\succcurlyeq$ High Influenza  & 3 & 35 & 8\\
			\hline
		\end{tabular} 
	}
	\label{tbl:rarePatterns}
	\vspace{-0.1in}
\end{table*}

%% file: rare_quantitativeEvaluation.tex
\input{experiment/rare_quantitativeBaselinesComparison}
\input{rare_tables_graphs}

\input{experiment/rare_quantitativeExactEvaluation}

%% file: experiment/rare_quantitativeBaselinesComparison.tex
\vspace{-0.1in}
\subsection{Baseline comparison on real world datasets}\label{sec:rare_baselines}\vspace{-0.05in}
We compare RTPM with the adapted baseline ARZ-Miner in terms of runtime and memory usage. 
Figs.  \ref{fig:rareruntimebaselineSC} and \ref{fig:rarememorybaselineSC} show the comparison results on SC.
As shown in Fig. \ref{fig:rareruntimebaselineSC}, RTPM has better runtime than the baseline. The range and average speedups of RTPM compared to the baseline is $[3.2$-$26.9]$ and $8.6$ on average. 
In terms of memory consumption, as shown in Fig. \ref{fig:rarememorybaselineSC}, RTPM uses less memory than the baseline. RTPM uses $[5.8$-$65.2]$ (on average $16.4$) times less memory than ARZ-Miner.

%% file: rare_tables_graphs.tex
\input{graph/rare_baselineComparison}
\input{graph/rare_runtimePruning}

%% file: graph/rare_baselineComparison.tex
\vspace{-0.1in}
\begin{figure*}[!t]
	\begin{minipage}[t]{1\columnwidth} 
		\centering
		\begin{subfigure}{0.32\columnwidth}
			\centering
			\resizebox{\linewidth}{!}{
				\begin{tikzpicture}[scale=0.2]
					\begin{axis}[
						compat=newest,
						xlabel={$\sigma_{\min}$ (\%)},
						ylabel={Runtime (sec)}, 
						label style={font=\Huge},
						ticklabel style = {font=\Huge},
						xmin=1, xmax=5,
						xtick={1,2,3,4,5},
						xticklabels = {1,3,6,9,12},
						ymin=1, ymax=100,
						legend columns=-1,
						legend entries = {RTPM, ARZ-Miner},
						legend style={nodes={scale=0.5,  transform shape}, font=\Large},
						legend to name={legendcomparisonrare},
						ymode=log,
						log basis y={10},
						ymajorgrids=true,
						grid style=dashed,
						line width=1.75pt
						]
						\addplot[
						color=teal,
						mark=diamond,
						mark size=4pt,
						]
						coordinates {
							(1,7)(2,4.7)(3,4.0)(4,2.0)(5,1.0)
						};
						\addplot[
						color=red,
						mark=pentagon*,
						mark size=4pt,
						]
						coordinates {
							(1,28)(2,23)(3,20)(4,15)(5,7)
						};
					\end{axis}
				\end{tikzpicture}
			}
			\captionsetup{justification=centering, font=scriptsize}
			\caption{\scriptsize Varying $\sigma_{\min}$}
		\end{subfigure}
		\begin{subfigure}{0.32\columnwidth}
			\centering
			\resizebox{\linewidth}{!}{
				\begin{tikzpicture}[scale=0.2]
					\begin{axis}[
						compat=newest,
						xlabel={$\delta$ (\%)},
						ylabel={Runtime (sec)}, 
						label style={font=\Huge},
						ticklabel style = {font=\Huge},
						xmin=60, xmax=100,
						ymin=1, ymax=100,
						xtick={60,70,80,90,100},
						legend columns=-1,
						legend entries = {RTPM, ARZ-Miner},
						legend style={nodes={scale=0.5,  transform shape}, font=\Large},
						legend to name={legendcomparisonrare},
						ymode=log,
						log basis y={10},
						ymajorgrids=true,
						grid style=dashed,
						line width=1.75pt
						]
						\addplot[
						color=teal,
						mark=diamond,
						mark size=4pt,
						]
						coordinates {
							(60,7)(70,5.5)(80,5.1)(90,3.5)(100,1.7)
						};
						\addplot[
						color=red,
						mark=pentagon*,
						mark size=4pt,
						]
						coordinates {
							(60,28)(70,25)(80,24)(90,18)(100,12)
						};
					\end{axis}
				\end{tikzpicture}
			}
			\captionsetup{justification=centering, font=scriptsize}
			\caption{\scriptsize Varying $\delta$}
		\end{subfigure}
		\begin{subfigure}{0.32\columnwidth}
			\centering
			\resizebox{\linewidth}{!}{
				\begin{tikzpicture}[scale=0.2]
					\begin{axis}[
						compat=newest,
						xlabel={$\sigma_{\max}$ (\%)},
						ylabel={Runtime (sec)}, 
						label style={font=\Huge},
						ticklabel style = {font=\Huge},
						xmin=1, xmax=5,
						ymin=1, ymax=100,
						xtick={1,2,3,4,5},
						xticklabels = {15,20,25,30,35},
						legend columns=-1,
						legend entries = {RTPM, ARZ-Miner},
						legend style={nodes={scale=0.5,  transform shape}, font=\Large},
						legend to name={legendcomparisonrare},
						ymode=log,
						log basis y={10},
						ymajorgrids=true,
						grid style=dashed,
						line width=1.75pt
						]
						\addplot[
						color=teal,
						mark=diamond,
						mark size=4pt,
						]
						coordinates {
							(1,4.6)(2,6.1)(3,6.5)(4,13)(5,15)
						};
						\addplot[
						color=red,
						mark=pentagon*,
						mark size=4pt,
						]
						coordinates {
							(1,21)(2,25)(3,28)(4,40)(5,46)
						};
					\end{axis}
				\end{tikzpicture}
			}
			\captionsetup{justification=centering, font=scriptsize}
			\caption{\scriptsize Varying $\sigma_{\max}$}
		\end{subfigure}
		\vspace{-0.1in}
		\captionsetup{justification=centering, font=small}
		\ref{legendcomparisonrare}
		\caption{Runtime Comparison on SC}
		\label{fig:rareruntimebaselineSC}
	\end{minipage}
\begin{minipage}[t]{1\columnwidth} 
	\centering
	\begin{subfigure}{0.32\columnwidth}
		\centering
		\resizebox{\linewidth}{!}{
			\begin{tikzpicture}[scale=0.2]
				\begin{axis}[
					compat=newest,
					xlabel={$\sigma_{\min}$ (\%)},
					ylabel={Memory Usage (MB)}, 
					label style={font=\Huge},
					ticklabel style = {font=\Huge},
					xmin=1, xmax=5,
					xtick={1,2,3,4,5},
					xticklabels = {1,3,6,9,12},
					ymin=10, ymax=1000,
					legend columns=-1,
					legend entries = {RTPM, ARZ-Miner},
					legend style={nodes={scale=0.5,  transform shape}, font=\Large},
					legend to name={legendcomparisonrare},
					ymode=log,
					log basis y={10},
					ymajorgrids=true,
					grid style=dashed,
					line width=1.75pt
					]
					\addplot[
					color=teal,
					mark=diamond,
					mark size=4pt,
					]
					coordinates {
						(1,101)(2,82)(3,65)(4,38)(5,26)
					};
					\addplot[
					color=red,
					mark=pentagon*,
					mark size=4pt,
					]
					coordinates {
						(1,428)(2,390)(3,302)(4,256)(5,177)
					};
				\end{axis}
			\end{tikzpicture}
		}
		\captionsetup{justification=centering, font=scriptsize}
		\caption{\scriptsize Varying $\sigma_{\min}$}
	\end{subfigure}
	\begin{subfigure}{0.32\columnwidth}
		\centering
		\resizebox{\linewidth}{!}{
			\begin{tikzpicture}[scale=0.2]
				\begin{axis}[
					compat=newest,
					xlabel={$\delta$ (\%)},
					ylabel={Memory Usage (MB)}, 
					label style={font=\Huge},
					ticklabel style = {font=\Huge},
					xmin=60, xmax=100,
					ymin=10, ymax=1000,
					xtick={60,70,80,90,100},
					legend columns=-1,
					legend entries = {RTPM, ARZ-Miner},
					legend style={nodes={scale=0.5,  transform shape}, font=\Large},
					legend to name={legendcomparisonrare},
					ymode=log,
					log basis y={10},
					ymajorgrids=true,
					grid style=dashed,
					line width=1.75pt
					]
					\addplot[
					color=teal,
					mark=diamond,
					mark size=4pt,
					]
					coordinates {
						(60,102)(70,84)(80,72)(90,54)(100,50)
					};
					\addplot[
					color=red,
					mark=pentagon*,
					mark size=4pt,
					]
					coordinates {
						(60,428)(70,406)(80,382)(90,361)(100,304)
					};
				\end{axis}
			\end{tikzpicture}
		}
		\captionsetup{justification=centering, font=scriptsize}
		\caption{\scriptsize Varying $\delta$}
	\end{subfigure}
	\begin{subfigure}{0.32\columnwidth}
		\centering
		\resizebox{\linewidth}{!}{
			\begin{tikzpicture}[scale=0.2]
				\begin{axis}[
					compat=newest,
					xlabel={$\sigma_{\max}$ (\%)},
					ylabel={Memory Usage (MB)}, 
					label style={font=\Huge},
					ticklabel style = {font=\Huge},
					xmin=1, xmax=5,
					ymin=10, ymax=1000,
					xtick={1,2,3,4,5},
					xticklabels = {15,20,25,30,35},
					legend columns=-1,
					legend entries = {RTPM, ARZ-Miner},
					legend style={nodes={scale=0.5,  transform shape}, font=\Large},
					legend to name={legendcomparisonrare},
					ymode=log,
					log basis y={10},
					ymajorgrids=true,
					grid style=dashed,
					line width=1.75pt
					]
					\addplot[
					color=teal,
					mark=diamond,
					mark size=4pt,
					]
					coordinates {
						(1,50)(2,76)(3,100)(4,120)(5,140)
					};
					\addplot[
					color=red,
					mark=pentagon*,
					mark size=4pt,
					]
					coordinates {
						(1,302)(2,358)(3,428)(4,542)(5,598)
					};
				\end{axis}
			\end{tikzpicture}
		}
		\captionsetup{justification=centering, font=scriptsize}
		\caption{\scriptsize Varying $\sigma_{\max}$}
	\end{subfigure}
	\vspace{-0.1in}
	\captionsetup{justification=centering, font=small}
	\ref{legendcomparisonrare}
	\caption{Memory Usage Comparison on SC}
	\label{fig:rarememorybaselineSC}
\end{minipage}
	\vspace{-0.1in}  
\end{figure*}   

%% file: graph/rare_runtimePruning.tex
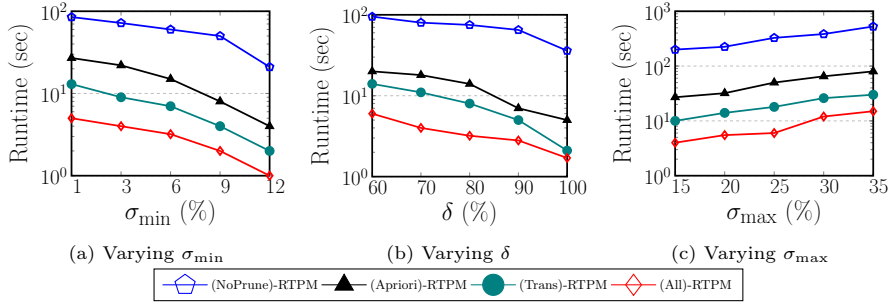
\begin{figure*}[!t]
	\begin{minipage}[t]{1\columnwidth} 
		\centering
		\begin{subfigure}{0.32\columnwidth}
			\centering
			\resizebox{\linewidth}{!}{
				\begin{tikzpicture}[scale=0.6]
					\begin{axis}[
						compat=newest,
						xlabel={$\sigma_{\min}$ (\%)},
						ylabel={Runtime (sec)}, 
						label style={font=\Huge},
						ticklabel style = {font=\huge},
						xticklabel style = {xshift=2mm}, 
						ymin=1, ymax=100,
						xmin=1, xmax=5,
						xtick={1,2,3,4,5},
						xticklabels = {1,3,6,9,12},
						legend columns=-1,
						legend entries = {(NoPrune)-RTPM, (Apriori)-RTPM, (Trans)-RTPM, (All)-RTPM},
						legend style={nodes={scale=0.55,  transform shape}, font=\small},
						legend to name={legendpruning},
						ymode=log,
						log basis y={10},
						ymajorgrids=true,
						grid style=dashed,
						line width=1.75pt
						]
						\addplot[
						color=blue,
						mark=pentagon,
						mark size=4pt,
						] 	
						coordinates {
							(1,85)(2,72)(3,60)(4,50)(5,21)
						};
						
						\addplot[
						color=black,
						mark=triangle*,
						mark size=4pt,
						]	
						coordinates {
							(1,27)(2,22)(3,15)(4,8)(5,4)
						};
						
						\addplot[
						color=teal,
						mark=*,
						mark size=4pt,
						] 
						coordinates {
							(1,13)(2,9)(3,7)(4,4)(5,2)
						};
						
						\addplot[
						color=red,
						mark=diamond,
						mark size=4pt,
						] 
						coordinates {
							(1,5)(2,4)(3,3.2)(4,2.0)(5,1)
						};
					\end{axis}
				\end{tikzpicture}
			}
			\captionsetup{justification=centering, font=scriptsize}
			\caption{Varying $\sigma_{\min}$}
		\end{subfigure}
		\begin{subfigure}{0.32\columnwidth}
			\centering
			\resizebox{\linewidth}{!}{
				\begin{tikzpicture}[scale=0.6]
					\begin{axis}[
						compat=newest,
						xlabel={$\delta$ (\%)},
						ylabel={Runtime (sec)}, 
						label style={font=\Huge},
						ticklabel style = {font=\huge},
						xticklabel style = {xshift=2mm},
						xmin=60, xmax=100,
						ymin=1, ymax=100,
						xtick={60,70,80,90,100},
						legend columns=-1,
						legend entries = {(NoPrune)-RTPM, (Apriori)-RTPM, (Trans)-RTPM, (All)-RTPM},
						legend style={nodes={scale=0.55,  transform shape}, font=\small},
						legend to name={legendpruning},
						ymode=log,
						log basis y={10},
						ymajorgrids=true,
						grid style=dashed,
						line width=1.75pt
						]
						\addplot[
						color=blue,
						mark=pentagon,
						mark size=4pt,
						] 	
						coordinates {
							(60,95)(70,80)(80,75)(90,65)(100,36)
						};
						
						\addplot[
						color=black,
						mark=triangle*,
						mark size=4pt,
						]	
						coordinates {
							(60,20)(70,18)(80,14)(90,7)(100,5)
						};
						
						\addplot[
						color=teal,
						mark=*,
						mark size=4pt,
						] 
						coordinates {
							(60,14)(70,11)(80,8)(90,5)(100,2.1)
						};
						
						\addplot[
						color=red,
						mark=diamond,
						mark size=4pt,
						] 
						coordinates {
							(60,6)(70,4)(80,3.2)(90,2.8)(100,1.7)
						};
					\end{axis}
				\end{tikzpicture}
			}
			\captionsetup{justification=centering, font=scriptsize}
			\caption{Varying $\delta$}
		\end{subfigure}
		\begin{subfigure}{0.32\columnwidth}
			\centering
			\resizebox{\linewidth}{!}{
				\begin{tikzpicture}[scale=0.6]
					\begin{axis}[
						compat=newest,
						xlabel={$\sigma_{\max}$ (\%)},
						ylabel={Runtime (sec)}, 
						label style={font=\Huge},
						ticklabel style = {font=\huge},
						xticklabel style = {xshift=2mm}, 
						xmin=1, xmax=5,
						ymin=1, ymax=1000,
						xtick={1,2,3,4,5},
						xticklabels = {15,20,25,30,35},
						legend columns=-1,
						legend entries = {(NoPrune)-RTPM, (Apriori)-RTPM, (Trans)-RTPM, (All)-RTPM},
						legend style={nodes={scale=0.55,  transform shape}, font=\small},
						legend to name={legendpruning},
						ymode=log,
						log basis y={10},
						ymajorgrids=true,
						grid style=dashed,
						line width=1.75pt
						]
						\addplot[
						color=blue,
						mark=pentagon,
						mark size=4pt,
						] 
						coordinates {
							(1,200)(2,225)(3,327)(4,384)(5,527)
						};
						
						\addplot[
						color=black,
						mark=triangle*,
						mark size=4pt,
						] 
						coordinates {
							(1,27)(2,32)(3,50)(4,65)(5,80)
						};
						
						\addplot[
						color=teal,
						mark=*,
						mark size=4pt,
						]	
						coordinates {
							(1,10)(2,14)(3,18)(4,26)(5,30)
						};
						
						\addplot[
						color=red,
						mark=diamond,
						mark size=4pt,
						] 
						coordinates {
							(1,4)(2,5.5)(3,6)(4,12)(5,15)
						};
					\end{axis}
				\end{tikzpicture}
			}
			\captionsetup{justification=centering, font=scriptsize}
			\caption{Varying $\sigma_{\max}$}
		\end{subfigure}
		\vspace{-0.1in}
		\ref{legendpruning}
		\caption{Runtimes of RTPM on SC} 
		\label{fig:rarepruningExactRTPM2}
	\end{minipage}      
	\vspace{-0.3in}  
\end{figure*}

%% file: experiment/rare_quantitativeExactEvaluation.tex
\vspace{-0.1in}
\subsection{Evaluation of different pruning techniques in RTPM}\label{rare_sec_exact}\vspace{-0.05in}
We evaluate the following combinations of RTPM pruning techniques: (1) NoPrune: RTPM with no pruning, (2) Apriori: RTPM with Apriori-based pruning (Lemmas \ref{lem2}, \ref{lem3}), (3) Trans: RTPM with transitivity-based pruning (Lemmas \ref{lem:transitivity}, \ref{lem:filter}, \ref{lem5}, \ref{lem6}), and (4) All: RTPM applied both pruning techniques. 

We use $3$ different scenarios that vary: the minimum support, the minimum confidence, and the maximum support. Fig.  \ref{fig:rarepruningExactRTPM2} 
shows the results. We see that (All)-RTPM has the best performance of all versions, with a speedup over (NoPrune)-RTPM ranging from $14$ up to $79$, depending on the configurations. Thus, the proposed prunings are very effective in improving RTPM performance. Furthermore, (Trans)-RTPM delivers a larger speedup than (Apriori)-RTPM, with the average speedup between $14$ and $29$ for (Trans)-RTPM, and between $5$ and $16$  for (Apriori)-RTPM, but applying both yields the best speedup.

%% file: Conclusion.tex
\vspace{-0.2in}
\section{Conclusion and Future Work}\label{sec:conclusion}\vspace{-0.1in}
This paper presents our efficient Rare Temporal Pattern Mining from Time Series (RTPMfTS) solution that offers: (1) an end-to-end RTPMfTS process to mine rare temporal patterns from time series, (2) an efficient Rare Temporal Pattern Mining (RTPM) algorithm that employs efficient data structures and multiple pruning techniques to achieve fast mining. Extensive experiments conducted on real world datasets show that our algorithm outperforms the baseline in runtime and memory usage. In future work, we plan to use the correlation measurement such as mutual information to prune uncorrelated time series, thereby further improve the performance.

%% file: Appendix.tex
\section{Appendix A: Detailed Proofs of Complexities, Lemmas}
\input{appendix/complexity_lemma_theorem}
\section{Appendix B: Additional Experimental Results}
\subsection{Baselines comparison on real world datasets}

Figs. \ref{fig:rareruntimebaselineINF}, \ref{fig:rareruntimebaselineNIST}, \ref{fig:rarememorybaselineINF}, and \ref{fig:rarememorybaselineNIST} show the comparison results among RTPM and the baseline on the INF and SE datasets. RTPM achieves the best performance (both runtime and memory usage) than the baseline. The speedup of RTPM compared to the baseline is [3.1-26.5] and 9.2 on average. 

On average, RTPM consumes ${\sim}5.3$ times less memory than the baseline.

\subsection{Evaluation of different pruning techniques in RTPM}
Figs. \ref{fig:rarepruningExactRTPM4_appendix} and \ref{fig:rarepruningExactRTPM1} 
show the results for the pruning techniques in RTPM on the INF and SE datasets. We see that (All)-RTPM has the best performance of all versions, with a speedup over (NoPrune)-RTPM ranging from $10$ up to $62$, depending on the configurations. Thus, the proposed prunings are very effective in improving RTPM performance. Furthermore, (Trans)-RTPM delivers a larger speedup than (Apriori)-RTPM, with the average speedup between $7$ and $24$ for (Trans)-RTPM, and between $5$ and $15$  for (Apriori)-E-RTPM, but applying both yields the best speedup.

\input{appendix/rare_baselineComparison}
\input{appendix/rare_runtimePruning}

%% file: appendix/complexity_lemma_theorem.tex
\subsection{Mutual exclusive property of temporal relations}\label{app:mutualexclusive}
\textbf{Property 1. }\textit{(Mutual exclusive) Consider the set of temporal relations $\Re=$ \{Follows, Contains, Overlaps\}. Let $E_i$ and $E_j$ be two temporal events, and $e_i$ occurring during $[t_{s_i}, t_{e_i}]$, $e_j$ occurring during $[t_{s_j}, t_{e_j}]$ be their corresponding event instances, and $\epsilon$ be the tolerance buffer. The relations in $\Re$ are mutually exclusive on $E_i$ and $E_j$.} 
\input{appendix/MutualExclusive}

\subsection{The search space of RTPM} 
\textit{Let $m$ be the number of distinct events in $\mathcal{D}_{\text{SEQ}}$, and $h$ be the longest length of a temporal pattern. The total number of temporal patterns is $O(m^h3^{h^2})$.}
\begin{proof}
	The number of single events is: $N_1=m \sim O(m)$. For mining event pairs, the number of permutations of $m$ distinct events taken $2$ at a time is: $P(m,2)$. However, since the same event can form a pair of events with itself, e.g., (SOn,SOn), the total number of event pairs is: $N_2=P(m,2)+m$ $\sim O(m^2)$. Each event pair in $N_2$ can form $3$ different temporal relations, and thus, the total number of 2-event patterns is: $N_2\times 3^1$ $\sim O(m^23^1)$. Similarly, the number of 3-event combinations is: $N_3= P(m,3)+P(m,2)+m$ $\sim O(m^3)$, and the number of 3-event patterns is: $N_3 \times 3^3$ $\sim O(m^33^3)$. For mining $h$-event combinations, the number of $h$-event combinations is $O(m^h)$, while the number of $h$-event patterns is $O(m^h \times 3^{\frac{1}{2}h(h-1)})$ $\sim O(m^h3^{h^2})$. Therefore, the total number of temporal patterns is $O(m)+O(m^23^1)+O(m^33^3)+...+O(m^h3^{h^2}) \sim O(m^h3^{h^2})$.
\end{proof}

\subsection{Lemma \ref{lem0}}
\textbf{Lemma \ref{lem0}. }\textit{Let $P$ be a temporal pattern and $E_i$ be a single event such that $E_i \in P$. Then $\textit{supp}(P) \le \textit{supp}(E_i)$.}
\begin{proof}
	Derived directly from Defs. 3.4, 3.5, and 3.8. 
\end{proof}

\subsection{Lemma \ref{lem2}} 
\textbf{Lemma \ref{lem2}. }\textit{Let $P$ be a 2-event pattern formed by an event pair $(E_i, E_j)$. Then, $\textit{supp}(P) \le \textit{supp}(E_i,E_j)$.}

From Lemma \ref{lem2}, the support of a pattern is at most the support of its events. Thus, infrequent event pairs (those which do not satisfy minimum support) cannot form frequent patterns and thereby, can be safely pruned.
\begin{proof}
	Derived directly from Defs. 3.4, 3.5, and 3.8. 
\end{proof}

\subsection{Lemma \ref{lem3}} 
\textbf{Lemma \ref{lem3}. }\textit{Let $(E_i, E_j)$ be a pair of events forming a 2-event pattern $P$.  Then \textit{conf}($P$) $\le$ \textit{conf}($E_i,E_j$).}

From Lemma \ref{lem3}, the confidence of a pattern $P$ is always at most the confidence of its events. Thus, a low-confidence event pair cannot form any high-confidence patterns and therefore, can be safely pruned.

\begin{proof}
	Can derived directly from Def. 3.9. 
\end{proof}

\subsection{Lemma \ref{lem:transitivity}} 
\textbf{Lemma \ref{lem:transitivity}. }\textit{Let $S=<e_1$,..., $e_{n-1}>$ be a temporal sequence that supports an (n-1)-event pattern $P=<(r_{12}$, $E_{1_{\triangleright e_1}}$, $E_{2_{\triangleright e_2}})$,..., $(r_{(n-2)(n-1)}$, $E_{{n-2}_{\triangleright e_{n-2}}}$, $E_{{n-1}_{\triangleright e_{n-1}}})>$. Let $e_n$ be a new event instance added to $S$ to create the temporal sequence $S^{'}$$=$$<e_1, ..., e_{n}>$.}
	
	\textit{The set of temporal relations $\Re$ is transitive on $S^{'}$: $\forall e_i \in S^{'}$, $i < n$, $\exists r \in \Re$ s.t. $r(E_{i_{\triangleright e_i}}$,$E_{n_{\triangleright e_n}})$ holds.}

Lemma \ref{lem:transitivity} says that given a temporal sequence $S$, a new event instance added to $S$ will always form at least one temporal relation with existing instances in $S$. This is due to the temporal transitivity property, as the time interval of new event instance will have a temporal order with the time intervals of existing instances.
\begin{proof}
	Since $S^{'}=<e_1, ..., e_{n}>$ is a temporal sequence, the event instances in $S^{'}$ are chronologically ordered by their start times. Then, $\forall e_i \in S^{'}, i \neq n$: $t_{s_i} \le t_{s_n}$. We have: 
	\begin{itemize}
		\item If $t_{e_i} \pm \epsilon \le t_{s_n}$, then $E_{i_{\triangleright e_i}}$ $\rightarrow E_{n_{\triangleright e_n}}$.
		\item If ${(t_{s_i} \le t_{s_n})} \wedge$ $(t_{e_i} \pm \epsilon \ge t_{e_n})$, then $E_{i_{\triangleright e_i}} \succcurlyeq E_{n_{\triangleright e_n}}$.
		\item If ${(t_{s_i} < t_{s_n})} \wedge$ $(t_{e_i} \pm \epsilon < t_{e_n})$ $\wedge$ $(t_{e_i}-t_{s_n} \ge d_o \pm \epsilon)$ where $d_o$ is the minimal overlapping duration, then $E_{i_{\triangleright e_i}} \between E_{n_{\triangleright e_n}}$. 
	\end{itemize}
\end{proof}

\subsection{Lemma \ref{lem:filter}} 
\textbf{Lemma \ref{lem:filter}. }\textit{Let $N_{k-1}=(E_1,...,E_{k-1})$ be a (\textit{k-1})-event combination and $E_k$ be a single event, both satisfying the $\sigma_{\min}$ constraint. The combination $N_k= N_{k-1} \cup E_k$ can form k-event temporal patterns whose support is at least $\sigma_{\min}$ if $\forall E_i \in N_{k-1}$, $\exists r \in \Re$ s.t. $r(E_i,E_k)$ is a frequent temporal relation.}

\begin{proof}
	Let $p_k$ be any k-event pattern formed by $N_k$. Then $p_k$ is a list of $\frac{1}{2}k(k-1)$ triples $(E_i,r_{ij},E_j)$ where each represents a relation $r(E_i,E_j)$ between two events. In order for $p_k$ to be frequent (satisfying $\sigma_{\min}$), each of the relations in $p_k$ must be frequent (Defs. 3.4 and 3.8, and Lemma \ref{lem:transitivity}).
\end{proof}

\subsection{Lemma \ref{lem5}}
\textbf{Lemma \ref{lem5}. }\textit{Let $P$ and $P^{'}$ be two temporal patterns. If $P^{'} \subseteq P$, then \textit{conf}($P^{'}$) $ \geq$ \textit{conf}($P$).}

Lemma \ref{lem5} says that the confidence of a pattern $P$ is always at most the confidence of its sub-patterns. 
\begin{proof}
	Can be derived directly from Def. 3.9.
\end{proof}

\subsection{Lemma \ref{lem6}} 
\textbf{Lemma \ref{lem6}. }\textit{Let $P$ and $P^{'}$ be two temporal patterns. If $P^{'} \subseteq P$ and \\ $\frac{\textit{supp}(P^{'})}{\max_{1 \leq k \leq \mid P \mid}\{\textit{supp}(E_k)\}}_{E_k \in P} \leq \delta$, then \textit{conf($P$)} $\leq \delta$.}

From Lemma \ref{lem6}, a temporal pattern $P$ cannot be high-confidence if any of its sub-patterns are low-confidence.
\begin{proof} We have:
	\begin{align}
		\small
		\text{\it{conf(P)}} &= \frac{\textit{supp}(P)}{\max_{1 \leq k \leq \mid P \mid}\{\textit{supp}(E_k) \}} \nonumber \\ &\leq \frac{\textit{supp}(P^{'})}{\max_{1 \leq k \leq \mid P \mid}\{\textit{supp}(E_k) \}} \leq \delta \nonumber
	\end{align}
\end{proof}

%% file: appendix/MutualExclusive.tex
\begin{proof}
	\textbf{$\ast$ Case 1:} Assume the relation \textbf{Follows($E_{i_{\triangleright e_i}}$, $E_{j_{\triangleright e_j}}$)} holds between $E_i$ and $E_j$. Thus, we have: 
	\begin{equation}
		t_{e_i} \pm \epsilon \le t_{s_j}
		\label{eq:mutualcase11}
	\end{equation}
	and:
	\begin{equation}
		t_{s_j} < t_{e_j} \Rightarrow t_{e_i} \pm \epsilon < t_{e_j}
		\label{eq:mutualcase12}
	\end{equation}
	
	Hence, Contains($E_{i_{\triangleright e_i}}, E_{j_{\triangleright e_j}}$) cannot exist between $E_i$ and $E_j$, since \\ Contains($E_{i_{\triangleright e_i}}$, $E_{j_{\triangleright e_j}}$) holds iff ${(t_{s_i} \le t_{s_j})} \wedge$ $(t_{e_i} \pm \epsilon \ge t_{e_j})$ (contradict Eq. \eqref{eq:mutualcase12}). 
	
	Similarly, Overlaps($E_{i_{\triangleright e_i}}, E_{j_{\triangleright e_j}}$) cannot exist between $E_i$ and $E_j$ since \\ {\small Overlaps}($E_{i_{\triangleright e_i}}$, $E_{j_{\triangleright e_j}}$) holds iff ${(t_{s_i} < t_{s_j})} \wedge$ $(t_{e_i} \pm \epsilon < t_{e_j})$ $\wedge$ $(t_{e_i}-t_{s_j} \ge d_o \pm \epsilon)$ (contradict Eq. \eqref{eq:mutualcase11}).
	
	In conclusion, if Follows($E_{i_{\triangleright e_i}}$, $E_{j_{\triangleright e_j}}$) holds between $E_i$ and $E_j$, then the two remaining relations cannot exist between $E_i$ and $E_j$.
	
	\textbf{$\ast$ Case 2:} Assume the relation \textbf{Contains($E_{i_{\triangleright e_i}}$, $E_{j_{\triangleright e_j}}$)} holds between $E_i$ and $E_j$. Thus, we have: 
	\begin{equation}
		t_{s_i} \le t_{s_j}
		\label{eq:mutualcase21}
	\end{equation}
	\begin{equation}
		t_{e_i} \pm \epsilon \ge t_{e_j}
		\label{eq:mutualcase22}
	\end{equation}
	
	Hence, Follows($E_{i_{\triangleright e_i}}$, $E_{j_{\triangleright e_j}}$) cannot exist between $E_i$ and $E_j$ since \\Follows($E_{i_{\triangleright e_i}}$, $E_{j_{\triangleright e_j}}$) holds iff $t_{e_i} \pm \epsilon < t_{e_j}$ (contradict Eq. \eqref{eq:mutualcase22}).
	
	Similarly, Overlaps($E_{i_{\triangleright e_i}}$, $E_{j_{\triangleright e_j}}$) cannot exist between $E_i$ and $E_j$, since \\Overlaps($E_{i_{\triangleright e_i}}$, $E_{j_{\triangleright e_j}}$) holds iff ${(t_{s_i} < t_{s_j})} \wedge$ $(t_{e_i} \pm \epsilon < t_{e_j})$ $\wedge$ $(t_{e_i}-t_{s_j} \ge d_o \pm \epsilon)$ (contradict Eq. \eqref{eq:mutualcase22}). 
	
	In conclusion, if Contains($E_{i_{\triangleright e_i}}$, $E_{j_{\triangleright e_j}}$) holds between $E_i$ and $E_j$, then the two remaining relations cannot exist between $E_i$ and $E_j$.
	
	\textbf{$\ast$ Case 3:} Assume the relation \textbf{Overlaps($E_{i_{\triangleright e_i}}$, $E_{j_{\triangleright e_j}}$)} holds between $E_i$ and $E_j$. Thus, we have: 
	\begin{equation}
		t_{s_i} < t_{s_j}
		\label{eq:mutualcase31}
	\end{equation}
	\begin{equation}
		t_{e_i} \pm \epsilon < t_{e_j}
		\label{eq:mutualcase32}
	\end{equation}
	\begin{equation}
		t_{e_i}-t_{s_j} \ge d_o \pm \epsilon \Rightarrow t_{s_j} \le t_{e_i} - d_o \pm \epsilon
		\label{eq:mutualcase33}
	\end{equation}
	
	Hence, Follows($E_{i_{\triangleright e_i}}$, $E_{j_{\triangleright e_j}}$) cannot exist between $E_i$ and $E_j$, since \\Follows($E_{i_{\triangleright e_i}}$, $E_{j_{\triangleright e_j}}$) holds iff $t_{e_i} \pm \epsilon < t_{s_j}$ (contradict Eq. \eqref{eq:mutualcase33}).
	
	Similarly, Contains($E_{i_{\triangleright e_i}}$, $E_{j_{\triangleright e_j}}$) cannot exist between $E_i$ and $E_j$, since \\Contains($E_{i_{\triangleright e_i}}$, $E_{j_{\triangleright e_j}}$) holds iff $t_{e_i} \pm \epsilon \ge t_{e_j}$ (contradict Eq. \eqref{eq:mutualcase32}).
	
	In conclusion, if Overlaps($E_{i_{\triangleright e_i}}$, $E_{j_{\triangleright e_j}}$) holds between $E_i$ and $E_j$, then the two remaining relations cannot exist between $E_i$ and $E_j$.
\end{proof}

%% file: appendix/rare_baselineComparison.tex
\begin{figure*}[!t]
	\begin{minipage}[t]{1\columnwidth} 
		\centering
		\begin{subfigure}{0.32\columnwidth}
			\centering
			\resizebox{\linewidth}{!}{
				\begin{tikzpicture}[scale=0.2]
					\begin{axis}[
						compat=newest,
						xlabel={$\sigma_{\min}$ (\%)},
						ylabel={Runtime (sec)}, 
						label style={font=\Huge},
						ticklabel style = {font=\Huge},
						xmin=1, xmax=5,
						xtick={1,2,3,4,5},
						xticklabels = {1,3,6,9,12},
						ymin=1, ymax=120,
						legend columns=-1,
						legend entries = {RTPM, ARZ-Miner},
						legend style={nodes={scale=0.5,  transform shape}, font=\Large},
						legend to name={legendcomparisonrare},
						ymode=log,
						log basis y={10},
						ymajorgrids=true,
						grid style=dashed,
						line width=1.75pt
						]
						\addplot[
						color=teal,
						mark=diamond,
						mark size=4pt,
						]
						coordinates {
							(1,32)(2,22)(3,10)(4,7)(5,5)
						};
						\addplot[
						color=red,
						mark=pentagon*,
						mark size=4pt,
						]
						coordinates {
							(1,95)(2,75)(3,62)(4,33)(5,14)
						};
					\end{axis}
				\end{tikzpicture}
			}
			\captionsetup{justification=centering, font=scriptsize}
			\caption{\scriptsize Varying $\sigma_{\min}$}
		\end{subfigure}
		\begin{subfigure}{0.32\columnwidth}
			\centering
			\resizebox{\linewidth}{!}{
				\begin{tikzpicture}[scale=0.2]
					\begin{axis}[
						compat=newest,
						xlabel={$\delta$ (\%)},
						ylabel={Runtime (sec)}, 
						label style={font=\Huge},
						ticklabel style = {font=\Huge},
						xmin=60, xmax=100,
						ymin=10, ymax=120,
						xtick={60,70,80,90,100},
						legend columns=-1,
						legend entries = {RTPM, ARZ-Miner},
						legend style={nodes={scale=0.5,  transform shape}, font=\Large},
						legend to name={legendcomparisonrare},
						ymode=log,
						log basis y={10},
						ymajorgrids=true,
						grid style=dashed,
						line width=1.75pt
						]
						\addplot[
						color=teal,
						mark=diamond,
						mark size=4pt,
						]
						coordinates {
							(60,32)(70,26)(80,21)(90,13)(100,10)
						};
						\addplot[
						color=red,
						mark=pentagon*,
						mark size=4pt,
						]
						coordinates {
							(60,95)(70,81)(80,72)(90,41)(100,20)
						};
					\end{axis}
				\end{tikzpicture}
			}
			\captionsetup{justification=centering, font=scriptsize}
			\caption{\scriptsize Varying $\delta$}
		\end{subfigure}
		\begin{subfigure}{0.32\columnwidth}
			\centering
			\resizebox{\linewidth}{!}{
				\begin{tikzpicture}[scale=0.2]
					\begin{axis}[
						compat=newest,
						xlabel={$\sigma_{\max}$ (\%)},
						ylabel={Runtime (sec)}, 
						label style={font=\Huge},
						ticklabel style = {font=\Huge},
						xmin=1, xmax=5,
						xtick={1,2,3,4,5},
						xticklabels = {15,20,25,30,35},
						ymin=10, ymax=120,
						legend columns=-1,
						legend entries = {RTPM, ARZ-Miner},
						legend style={nodes={scale=0.5,  transform shape}, font=\Large},
						legend to name={legendcomparisonrare},
						ymode=log,
						log basis y={10},
						ymajorgrids=true,
						grid style=dashed,
						line width=1.75pt
						]
						\addplot[
						color=teal,
						mark=diamond,
						mark size=4pt,
						]
						coordinates {
							(1,12)(2,25)(3,32)(4,37)(5,48)
						};
						\addplot[
						color=red,
						mark=pentagon*,
						mark size=4pt,
						]
						coordinates {
							(1,51)(2,74)(3,95)(4,105)(5,119)
						};
					\end{axis}
				\end{tikzpicture}
			}
			\captionsetup{justification=centering, font=scriptsize}
			\caption{\scriptsize Varying $\sigma_{\max}$}
		\end{subfigure}
		\captionsetup{justification=centering, font=small}
		\ref{legendcomparisonrare}
		\caption{Runtime Comparison on INF}
		\label{fig:rareruntimebaselineINF}
	\end{minipage}  
\begin{minipage}[t]{1\columnwidth} 
	\centering
	\begin{subfigure}{0.32\columnwidth}
		\centering
		\resizebox{\linewidth}{!}{
			\begin{tikzpicture}[scale=0.2]
				\begin{axis}[
					compat=newest,
					xlabel={$\sigma_{\min}$ (\%)},
					ylabel={Runtime (sec)}, 
					label style={font=\Huge},
					ticklabel style = {font=\Huge},
					xmin=1, xmax=5,
					xtick={1,2,3,4,5},
					xticklabels = {1,3,6,9,12},
					ymin=10, ymax=10000,
					legend columns=-1,
					legend entries = {RTPM, ARZ-Miner},
					legend style={nodes={scale=0.5,  transform shape}, font=\Large},
					legend to name={legendcomparisonrare},
					ymode=log,
					log basis y={10},
					ymajorgrids=true,
					grid style=dashed,
					line width=1.75pt
					]						
					\addplot[
					color=teal,
					mark=diamond,
					mark size=4pt,
					]
					coordinates {
						(1,340)(2,187)(3,120)(4,93)(5,77)
					};
					\addplot[
					color=red,
					mark=pentagon*,
					mark size=4pt,
					]
					coordinates {
						(1,1502)(2,926)(3,734)(4,616)(5,452)
					};
				\end{axis}
			\end{tikzpicture}
		}
		\captionsetup{ font=scriptsize}
		\caption{\scriptsize Varying $\sigma_{\min}$}
	\end{subfigure}
	\begin{subfigure}{0.32\columnwidth}
		\centering
		\resizebox{\linewidth}{!}{
			\begin{tikzpicture}[scale=0.2]
				\begin{axis}[
					compat=newest,
					xlabel={$\delta$ (\%)},
					ylabel={Runtime (sec)}, 
					label style={font=\Huge},
					ticklabel style = {font=\Huge},
					xmin=60, xmax=100,
					ymin=1, ymax=10000,
					xtick={60,70,80,90,100},
					legend columns=-1,
					legend entries = {RTPM, ARZ-Miner},
					legend style={nodes={scale=0.5,  transform shape}, font=\Large},
					legend to name={legendcomparisonrare},
					ymode=log,
					log basis y={10},
					ymajorgrids=true,
					grid style=dashed,
					line width=1.75pt
					]
					\addplot[
					color=teal,
					mark=diamond,
					mark size=4pt,
					]
					coordinates {
						(60,340)(70,267)(80,164)(90,50)(100,1.5)
					};
					\addplot[
					color=red,
					mark=pentagon*,
					mark size=4pt,
					]
					coordinates {
						(60,1502)(70,1225)(80,909)(90,461)(100,10)
					};
				\end{axis}
			\end{tikzpicture}
		}
		\captionsetup{justification=centering, font=scriptsize}
		\caption{\scriptsize Varying $\delta$}
	\end{subfigure}
	\begin{subfigure}{0.32\columnwidth}
		\centering
		\resizebox{\linewidth}{!}{
			\begin{tikzpicture}[scale=0.2]
				\begin{axis}[
					compat=newest,
					xlabel={$\sigma_{\max}$ (\%)},
					ylabel={Runtime (sec)}, 
					label style={font=\Huge},
					ticklabel style = {font=\Huge},
					xmin=1, xmax=5,
					ymin=100, ymax=10000,
					xtick={1,2,3,4,5},
					xticklabels = {15,20,25,30,35},
					legend columns=-1,
					legend entries = {RTPM, ARZ-Miner},
					legend style={nodes={scale=0.5,  transform shape}, font=\Large},
					legend to name={legendcomparisonrare},
					ymode=log,
					log basis y={10},
					ymajorgrids=true,
					grid style=dashed,
					line width=1.75pt
					]
					\addplot[
					color=teal,
					mark=diamond,
					mark size=4pt,
					]
					coordinates {
						(1,194)(2,218)(3,340)(4,382)(5,401)
					};
					\addplot[
					color=red,
					mark=pentagon*,
					mark size=4pt,
					]
					coordinates {
						(1,926)(2,1024)(3,1902)(4,2802)(5,3967)
					};
				\end{axis}
			\end{tikzpicture}
		}
		\captionsetup{justification=centering, font=scriptsize}
		\caption{\scriptsize Varying $\sigma_{\max}$}
	\end{subfigure}
	\vspace{-0.1in}
	\captionsetup{justification=centering, font=small}
	\ref{legendcomparisonrare}
	\caption{Runtime Comparison on SE}
	\label{fig:rareruntimebaselineNIST}
\end{minipage}
\end{figure*}   

\begin{figure*}[!t]
	\begin{minipage}[t]{1\columnwidth} 
		\centering
		\begin{subfigure}{0.32\columnwidth}
			\centering
			\resizebox{\linewidth}{!}{
				\begin{tikzpicture}[scale=0.2]
					\begin{axis}[
						compat=newest,
						xlabel={$\sigma_{\min}$ (\%)},
						ylabel={Memory Usage (MB)}, 
						label style={font=\Huge},
						ticklabel style = {font=\Huge},
						xmin=1, xmax=5,
						xtick={1,2,3,4,5},
						xticklabels = {1,3,6,9,12},
						ymin=10, ymax=1000,
						legend columns=-1,
						legend entries = {RTPM, ARZ-Miner},
						legend style={nodes={scale=0.5,  transform shape}, font=\Large},
						legend to name={legendcomparisonrare},
						ymode=log,
						log basis y={10},
						ymajorgrids=true,
						grid style=dashed,
						line width=1.75pt
						]
						\addplot[
						color=teal,
						mark=diamond,
						mark size=4pt,
						]
						coordinates {
							(1,80)(2,70)(3,62)(4,51)(5,50)
						};
						\addplot[
						color=red,
						mark=pentagon*,
						mark size=4pt,
						]
						coordinates {
							(1,450)(2,352)(3,321)(4,258)(5,181)
						};
					\end{axis}
				\end{tikzpicture}
			}
			\captionsetup{justification=centering, font=scriptsize}
			\caption{\scriptsize Varying $\sigma_{\min}$}
		\end{subfigure}
		\begin{subfigure}{0.32\columnwidth}
			\centering
			\resizebox{\linewidth}{!}{
				\begin{tikzpicture}[scale=0.2]
					\begin{axis}[
						compat=newest,
						xlabel={$\delta$ (\%)},
						ylabel={Memory Usage (MB)}, 
						label style={font=\Huge},
						ticklabel style = {font=\Huge},
						xmin=60, xmax=100,
						ymin=10, ymax=1000,
						xtick={60,70,80,90,100},
						legend columns=-1,
						legend entries = {RTPM, ARZ-Miner},
						legend style={nodes={scale=0.5,  transform shape}, font=\Large},
						legend to name={legendcomparisonrare},
						ymode=log,
						log basis y={10},
						ymajorgrids=true,
						grid style=dashed,
						line width=1.75pt
						]
						\addplot[
						color=teal,
						mark=diamond,
						mark size=4pt,
						]
						coordinates {
							(60,80)(70,72)(80,65)(90,61)(100,52)
						};
						\addplot[
						color=red,
						mark=pentagon*,
						mark size=4pt,
						]
						coordinates {
							(60,450)(70,420)(80,372)(90,346)(100,285)
						};
					\end{axis}
				\end{tikzpicture}
			}
			\captionsetup{justification=centering, font=scriptsize}
			\caption{\scriptsize Varying $\delta$}
		\end{subfigure}
		\begin{subfigure}{0.32\columnwidth}
			\centering
			\resizebox{\linewidth}{!}{
				\begin{tikzpicture}[scale=0.2]
					\begin{axis}[
						compat=newest,
						xlabel={$\sigma_{\max}$ (\%)},
						ylabel={Memory Usage (MB)}, 
						label style={font=\Huge},
						ticklabel style = {font=\Huge},
						xmin=1, xmax=5,
						xtick={1,2,3,4,5},
						xticklabels = {15,20,25,30,35},
						ymin=10, ymax=1000,
						legend columns=-1,
						legend entries = {RTPM, ARZ-Miner},
						legend style={nodes={scale=0.5,  transform shape}, font=\Large},
						legend to name={legendcomparisonrare},
						ymode=log,
						log basis y={10},
						ymajorgrids=true,
						grid style=dashed,
						line width=1.75pt
						]
						\addplot[
						color=teal,
						mark=diamond,
						mark size=4pt,
						]
						coordinates {
							(1,56)(2,70)(3,80)(4,86)(5,95)
						};
						\addplot[
						color=red,
						mark=pentagon*,
						mark size=4pt,
						]
						coordinates {
							(1,358)(2,416)(3,450)(4,591)(5,796)
						};
					\end{axis}
				\end{tikzpicture}
			}
			\captionsetup{justification=centering, font=scriptsize}
			\caption{\scriptsize Varying $\sigma_{\max}$}
		\end{subfigure}
		\captionsetup{justification=centering, font=small}
		\ref{legendcomparisonrare}
		\caption{Memory Usage Comparison on INF}
		\label{fig:rarememorybaselineINF}
	\end{minipage}
\end{figure*}
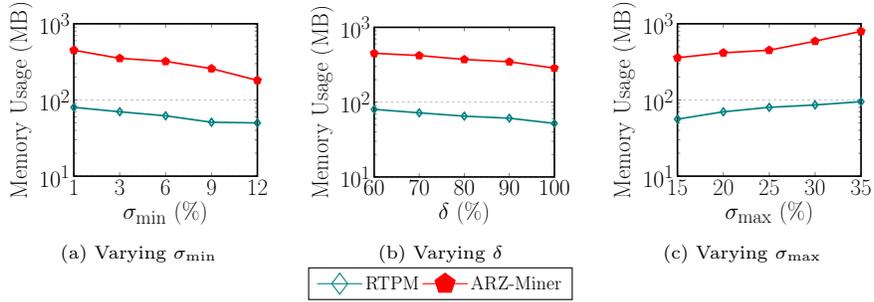  
\begin{figure*}[h]
	\begin{minipage}[t]{1\columnwidth} 
		\centering
		\begin{subfigure}{0.32\columnwidth}
			\centering
			\resizebox{\linewidth}{!}{
				\begin{tikzpicture}[scale=0.2]
					\begin{axis}[
						compat=newest,
						xlabel={$\sigma_{\min}$ (\%)},
						ylabel={Memory Usage (MB)}, 
						label style={font=\Huge},
						ticklabel style = {font=\Huge},
						xmin=1, xmax=5,
						xtick={1,2,3,4,5},
						xticklabels = {1,3,6,9,12},
						ymin=100, ymax=50000,
						legend columns=-1,
						legend entries = {RTPM, ARZ-Miner},
						legend style={nodes={scale=0.5,  transform shape}, font=\Large},
						legend to name={legendcomparisonrare},
						ymode=log,
						log basis y={10},
						ymajorgrids=true,
						grid style=dashed,
						line width=1.75pt
						]
						\addplot[
						color=teal,
						mark=diamond,
						mark size=4pt,
						]
						coordinates {
							(1,747)(2,581)(3,507)(4,364)(5,307)
						};
						\addplot[
						color=red,
						mark=pentagon*,
						mark size=4pt,
						]
						coordinates {
							(1,27206)(2,12620)(3,7684)(4,2708)(5,1706)
						};
					\end{axis}
				\end{tikzpicture}
			}
			\captionsetup{justification=centering, font=scriptsize}
			\caption{\scriptsize Varying $\sigma_{\min}$}
		\end{subfigure}
		\begin{subfigure}{0.32\columnwidth}
			\centering
			\resizebox{\linewidth}{!}{
				\begin{tikzpicture}[scale=0.2]
					\begin{axis}[
						compat=newest,
						xlabel={$\delta$ (\%)},
						ylabel={Memory Usage (MB)}, 
						label style={font=\Huge},
						ticklabel style = {font=\Huge},
						xmin=60, xmax=100,
						ymin=100, ymax=50000,
						xtick={60,70,80,90,100},
						legend columns=-1,
						legend entries = {RTPM, ARZ-Miner},
						legend style={nodes={scale=0.5,  transform shape}, font=\Large},
						legend to name={legendcomparisonrare},
						ymode=log,
						log basis y={10},
						ymajorgrids=true,
						grid style=dashed,
						line width=1.75pt
						]
						\addplot[
						color=teal,
						mark=diamond,
						mark size=4pt,
						]
						coordinates {
							(60,747)(70,614)(80,558)(90,376)(100,184)
						};
						\addplot[
						color=red,
						mark=pentagon*,
						mark size=4pt,
						]
						coordinates {
							(60,27206)(70,16247)(80,9006)(90,3216)(100,1026)
						};
					\end{axis}
				\end{tikzpicture}
			}
			\captionsetup{justification=centering, font=scriptsize}
			\caption{\scriptsize Varying $\delta$}
		\end{subfigure}
		\begin{subfigure}{0.32\columnwidth}
			\centering
			\resizebox{\linewidth}{!}{
				\begin{tikzpicture}[scale=0.2]
					\begin{axis}[
						compat=newest,
						xlabel={$\sigma_{\max}$ (\%)},
						ylabel={Memory Usage (MB)}, 
						label style={font=\Huge},
						ticklabel style = {font=\Huge},
						xmin=1, xmax=5,
						ymin=100, ymax=50000,
						xtick={1,2,3,4,5},
						xticklabels = {15,20,25,30,35},
						legend columns=-1,
						legend entries = {RTPM, ARZ-Miner},
						legend style={nodes={scale=0.5,  transform shape}, font=\Large},
						legend to name={legendcomparisonrare},
						ymode=log,
						log basis y={10},
						ymajorgrids=true,
						grid style=dashed,
						line width=1.75pt
						]
						\addplot[
						color=teal,
						mark=diamond,
						mark size=4pt,
						]
						coordinates {
							(1,452)(2,506)(3,747)(4,852)(5,957)
						};
						\addplot[
						color=red,
						mark=pentagon*,
						mark size=4pt,
						]
						coordinates {
							(1,4247)(2,6004)(3,27206)(4,39851)(5,48028)
						};
					\end{axis}
				\end{tikzpicture}
			}
			\captionsetup{justification=centering, font=scriptsize}
			\caption{\scriptsize Varying $\sigma_{\max}$}
		\end{subfigure}
		\vspace{-0.1in}
		\captionsetup{justification=centering, font=small}
		\ref{legendcomparisonrare}
		\caption{Memory Usage Comparison on  SE}
		\label{fig:rarememorybaselineNIST}
	\end{minipage}
\end{figure*}
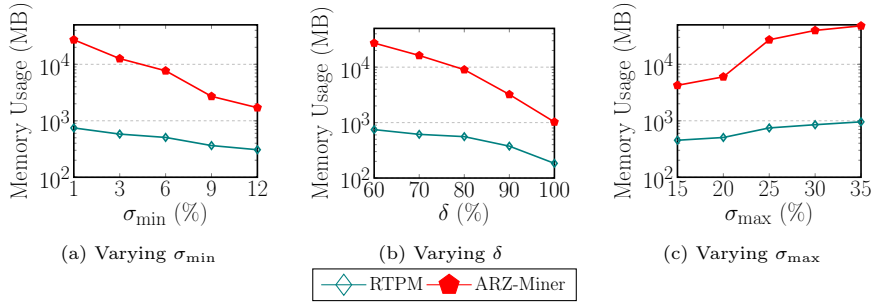

%% file: appendix/rare_runtimePruning.tex
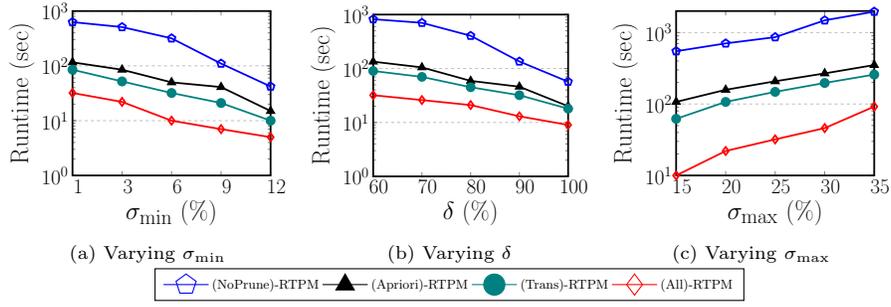
\begin{figure}[h]
	\begin{minipage}[t]{1\columnwidth} 
		\centering
		\begin{subfigure}{0.32\columnwidth}
			\centering
			\resizebox{\linewidth}{!}{
				\begin{tikzpicture}[scale=0.6]
					\begin{axis}[
						compat=newest,
						xlabel={$\sigma_{\min}$ (\%)},
						ylabel={Runtime (sec)}, 
						label style={font=\Huge},
						ticklabel style = {font=\huge},
						xticklabel style = {xshift=2mm}, 
						ymin=1, ymax=1000,
						xmin=1, xmax=5,
						xtick={1,2,3,4,5},
						xticklabels = {1,3,6,9,12},
						legend columns=-1,
						legend entries = {(NoPrune)-RTPM, (Apriori)-RTPM, (Trans)-RTPM, (All)-RTPM},
						legend style={nodes={scale=0.55,  transform shape}, font=\small},
						legend to name={legendpruning},
						ymode=log,
						log basis y={10},
						ymajorgrids=true,
						grid style=dashed,
						line width=1.75pt
						]
						\addplot[
						color=blue,
						mark=pentagon,
						mark size=4pt,
						] 	
						coordinates {
							(1,628)(2,510)(3,318)(4,110)(5,42)
						};
						
						\addplot[
						color=black,
						mark=triangle*,
						mark size=4pt,
						]	
						coordinates {
							(1,116)(2,85)(3,50)(4,41)(5,15)
						};
						
						\addplot[
						color=teal,
						mark=*,
						mark size=4pt,
						] 
						coordinates {
							(1,85)(2,52)(3,32)(4,21)(5,10)
						};
						
						\addplot[
						color=red,
						mark=diamond,
						mark size=4pt,
						] 
						coordinates {
							(1,32)(2,22)(3,10)(4,7)(5,5)
						};
					\end{axis}
				\end{tikzpicture}
			}
			\captionsetup{justification=centering, font=scriptsize}
			\caption{Varying $\sigma_{\min}$}
		\end{subfigure}
		\begin{subfigure}{0.32\columnwidth}
			\centering
			\resizebox{\linewidth}{!}{
				\begin{tikzpicture}[scale=0.6]
					\begin{axis}[
						compat=newest,
						xlabel={$\delta$ (\%)},
						ylabel={Runtime (sec)}, 
						label style={font=\Huge},
						ticklabel style = {font=\huge},
						xticklabel style = {xshift=2mm}, 
						xmin=60, xmax=100,
						ymin=1, ymax=1000,
						xtick={60,70,80,90,100},
						legend columns=-1,
						legend entries = {(NoPrune)-RTPM, (Apriori)-RTPM, (Trans)-RTPM, (All)-RTPM},
						legend style={nodes={scale=0.55,  transform shape}, font=\small},
						legend to name={legendpruning},
						ymode=log,
						log basis y={10},
						ymajorgrids=true,
						grid style=dashed,
						line width=1.75pt
						]
						\addplot[
						color=blue,
						mark=pentagon,
						mark size=4pt,
						] 	
						coordinates {
							(60,821)(70,704)(80,405)(90,135)(100,57)
						};
						
						\addplot[
						color=black,
						mark=triangle*,
						mark size=4pt,
						]	
						coordinates {
							(60,134)(70,105)(80,59)(90,46)(100,20)
						};
						
						\addplot[
						color=teal,
						mark=*,
						mark size=4pt,
						] 
						coordinates {
							(60,90)(70,70)(80,45)(90,32)(100,18)
						};
						
						\addplot[
						color=red,
						mark=diamond,
						mark size=4pt,
						] 
						coordinates {
							(60,32)(70,26)(80,21)(90,13)(100,9)
						};
					\end{axis}
				\end{tikzpicture}
			}
			\captionsetup{justification=centering, font=scriptsize}
			\caption{Varying $\delta$}
		\end{subfigure}
		\begin{subfigure}{0.32\columnwidth}
			\centering
			\resizebox{\linewidth}{!}{
				\begin{tikzpicture}[scale=0.6]
					\begin{axis}[
						compat=newest,
						xlabel={$\sigma_{\max}$ (\%)},
						ylabel={Runtime (sec)}, 
						label style={font=\Huge},
						ticklabel style = {font=\huge},
						xticklabel style = {xshift=2mm}, 
						xmin=1, xmax=5,
						xtick={1,2,3,4,5},
						xticklabels = {15,20,25,30,35},
						ymin=10, ymax=2000,
						legend columns=-1,
						legend entries = {(NoPrune)-RTPM, (Apriori)-RTPM, (Trans)-RTPM, (All)-RTPM},
						legend style={nodes={scale=0.55,  transform shape}, font=\small},
						legend to name={legendpruning},
						ymode=log,
						log basis y={10},
						ymajorgrids=true,
						grid style=dashed,
						line width=1.75pt
						]
						\addplot[
						color=blue,
						mark=pentagon,
						mark size=4pt,
						] 
						coordinates {
							(1,548)(2,705)(3,862)(4,1482)(5,1962)
						};
						
						\addplot[
						color=black,
						mark=triangle*,
						mark size=4pt,
						] 
						coordinates {
							(1,107)(2,158)(3,208)(4,268)(5,351)
						};
						
						\addplot[
						color=teal,
						mark=*,
						mark size=4pt,
						]	
						coordinates {
							(1,62)(2,107)(3,148)(4,197)(5,258)
						};
						
						\addplot[
						color=red,
						mark=diamond,
						mark size=4pt,
						] 
						coordinates {
							(1,10)(2,22)(3,32)(4,46)(5,92)
						};
					\end{axis}
				\end{tikzpicture}
			}
			\captionsetup{justification=centering, font=scriptsize}
			\caption{Varying $\sigma_{\max}$}
		\end{subfigure}
		\vspace{-0.1in}
		\ref{legendpruning}
		\caption{Runtimes of RTPM on INF} 
		\label{fig:rarepruningExactRTPM4_appendix}
	\end{minipage}      
\end{figure}  
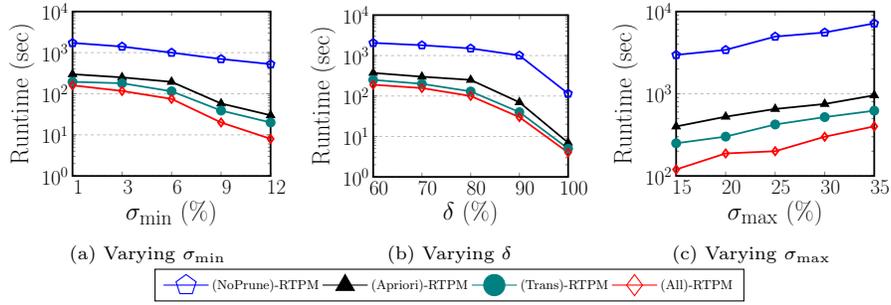
\begin{figure}[!h]
	\vspace{-3in}
	\begin{minipage}[t]{1\columnwidth} 
		\centering
		\begin{subfigure}{0.32\columnwidth}
			\centering
			\resizebox{\linewidth}{!}{
				\begin{tikzpicture}[scale=0.6]
					\begin{axis}[
						compat=newest,
						xlabel={$\sigma_{\min}$ (\%)},
						ylabel={Runtime (sec)}, 
						label style={font=\Huge},
						ticklabel style = {font=\huge},
						xticklabel style = {xshift=2mm}, 
						ymin=1, ymax=10000,
						xmin=1, xmax=5,
						xtick={1,2,3,4,5},
						xticklabels = {1,3,6,9,12},
						legend columns=-1,
						legend entries = {(NoPrune)-RTPM, (Apriori)-RTPM, (Trans)-RTPM, (All)-RTPM},
						legend style={nodes={scale=0.55,  transform shape}},
						legend to name={legendpruning},
						ymode=log,
						log basis y={10},
						ymajorgrids=true,
						grid style=dashed,
						line width=1.75pt
						]
						\addplot[
						color=blue,
						mark=pentagon,
						mark size=4pt,
						] 	
						coordinates {
							(1,1724)(2,1405)(3,1000)(4,700)(5,525)
						};
						
						\addplot[
						color=black,
						mark=triangle*,
						mark size=4pt,
						]	
						coordinates {
							(1,299)(2,250)(3,195)(4,58)(5,30)
						};
						
						\addplot[
						color=teal,
						mark=*,
						mark size=4pt,
						] 
						coordinates {
							(1,195)(2,180)(3,115)(4,39)(5,20)
						};
						
						\addplot[
						color=red,
						mark=diamond,
						mark size=4pt,
						] 
						coordinates {
							(1,160)(2,117)(3,75)(4,20)(5,8)
						};
					\end{axis}
				\end{tikzpicture}
			}
			\captionsetup{justification=centering, font=scriptsize}
			\caption{Varying $\sigma_{\min}$}
		\end{subfigure}
		\begin{subfigure}{0.32\columnwidth}
			\centering
			\resizebox{\linewidth}{!}{
				\begin{tikzpicture}[scale=0.6]
					\begin{axis}[
						compat=newest,
						xlabel={$\delta$ (\%)},
						ylabel={Runtime (sec)}, 
						label style={font=\Huge},
						ticklabel style = {font=\huge},
						xticklabel style = {xshift=2mm}, 
						xmin=60, xmax=100,
						ymin=1, ymax=10000,
						xtick={60,70,80,90,100},
						legend columns=-1,
						legend entries = {(NoPrune)-RTPM, (Apriori)-RTPM, (Trans)-RTPM, (All)-RTPM},
						legend style={nodes={scale=0.55,  transform shape}},
						legend to name={legendpruning},
						ymode=log,
						log basis y={10},
						ymajorgrids=true,
						grid style=dashed,
						line width=1.75pt
						]
						\addplot[
						color=blue,
						mark=pentagon,
						mark size=4pt,
						] 	
						coordinates {
							(60,2054)(70,1800)(80,1500)(90,1005)(100,115)
						};
						
						\addplot[
						color=black,
						mark=triangle*,
						mark size=4pt,
						]	
						coordinates {
							(60,372)(70,300)(80,250)(90,70)(100,7)
						};
						
						\addplot[
						color=teal,
						mark=*,
						mark size=4pt,
						] 
						coordinates {
							(60,250)(70,200)(80,130)(90,40)(100,5)
						};
						
						\addplot[
						color=red,
						mark=diamond,
						mark size=4pt,
						] 
						coordinates {
							(60,190)(70,157)(80,100)(90,30)(100,4)
						};
					\end{axis}
				\end{tikzpicture}
			}
			\captionsetup{justification=centering, font=scriptsize}
			\caption{Varying $\delta$}
		\end{subfigure}
		\begin{subfigure}{0.32\columnwidth}
			\centering
			\resizebox{\linewidth}{!}{
				\begin{tikzpicture}[scale=0.6]
					\begin{axis}[
						compat=newest,
						xlabel={$\sigma_{\max}$ (\%)},
						ylabel={Runtime (sec)}, 
						label style={font=\Huge},
						ticklabel style = {font=\huge},
						xticklabel style = {xshift=2mm}, 
						xmin=1, xmax=5,
						ymin=100, ymax=10000,
						xtick={1,2,3,4,5},
						xticklabels = {15,20,25,30,35},
						legend columns=-1,
						legend entries = {(NoPrune)-RTPM, (Apriori)-RTPM, (Trans)-RTPM, (All)-RTPM},
						legend style={nodes={scale=0.55,  transform shape}},
						legend to name={legendpruning},
						ymode=log,
						log basis y={10},
						ymajorgrids=true,
						grid style=dashed,
						line width=1.75pt
						]
						\addplot[
						color=blue,
						mark=pentagon,
						mark size=4pt,
						] 
						coordinates {
							(1,2969)(2,3416)(3,4958)(4,5561)(5,7169)
						};
						
						\addplot[
						color=black,
						mark=triangle*,
						mark size=4pt,
						] 
						coordinates {
							(1,400)(2,526)(3,653)(4,751)(5,958)
						};
						
						\addplot[
						color=teal,
						mark=*,
						mark size=4pt,
						]	
						coordinates {
							(1,250)(2,301)(3,423)(4,520)(5,622)
						};
						
						\addplot[
						color=red,
						mark=diamond,
						mark size=4pt,
						] 
						coordinates {
							(1,120)(2,188)(3,200)(4,300)(5,401)
						};
					\end{axis}
				\end{tikzpicture}
			}
			\captionsetup{justification=centering, font=scriptsize}
			\caption{Varying $\sigma_{\max}$}
		\end{subfigure}
		\vspace{-0.1in}
		\ref{legendpruning}
		\caption{Runtimes of RTPM on SE} 
		\label{fig:rarepruningExactRTPM1}
	\end{minipage}%
\end{figure}